\documentclass[12pt,amsmath,amsfonts,array,pstricks,color,calc,cite,axodraw,feynmp]{article}
\usepackage{amsmath}
\usepackage{amsfonts}
\usepackage{array}
\usepackage{pstricks}
\usepackage{color}
\usepackage{calc}
\usepackage{cite}
\usepackage{axodraw}
\usepackage{feynmp}

\newcommand{\myfigcaption}[2]{
  \refstepcounter{figure}
  \addtocontents{lof}{
    \protect\contentsline {figure}{\numberline {\thefigure}
      {\ignorespaces \string\BeforeSubString{\newline}{#2}}}{\thepage}
  }
  \begin{center}
    \small{\bf Fig.~\thefigure:\hspace*{2ex}}
    \parbox[t]{#1-\widthof{\bf Abb. \thefigure\hspace*{4ex}}}{#2}
  \end{center}
}
\newcommand{\myfigure}[3]{
  \begin{figure}[#1]
    \begin{center}#2\end{center}
    \myfigcaption{0.9\textwidth}{#3}
  \end{figure}
}
\newcommand{\mytabcaption}[2]{
  \refstepcounter{table}
  \addtocontents{lot}{
    \protect\contentsline {table}{\numberline {\thetable}
      {\ignorespaces \string\BeforeSubString{\newline}{#2}}}{\thepage}
  }
  \begin{center}
    \small{\bf Tab.~\thetable:\hspace*{2ex}}
    \parbox[t]{#1-\widthof{\bf Tab. \thetable\hspace*{4ex}}}{#2}
  \end{center}
}
\newcommand{\mytable}[3]{
  \begin{table}[#1]
    \begin{center}#2\end{center}
    \mytabcaption{0.9\textwidth}{#3}
  \end{table}
}
\newskip\humongous \humongous=0pt plus 1000pt minus 100pt

\newif\ifdtup

\newcounter{eqnumber}[section]

\makeatletter
\def\@eqnnum{\hbox{\reset@font\rm(\theequation)}}
\let\make@eqnnum=\@eqnnum %
\def\eqnum#1{\dec@eqnnum \global\def\make@eqnnum{\reset@font\rm(#1)}%
\def\@currentlabel{#1}%
}
\def\inc@eqnnum{\addtocounter{equation}{1}}
\def\dec@eqnnum{\addtocounter{equation}{-1}}
\@definecounter{equation}%
\@addtoreset{equation}{section} %
\def\theequation@prefix{{\thesection}.} %
\def\theequation{\theequation@prefix\arabic{equation}}%
\makeatother

\def\Sec#1{Section~{\ref{#1}}}

\catcode`@=11  
\long\def\symbolfootnote[#1]#2{\begingroup%
\def\thefootnote{\fnsymbol{footnote}}\footnote[#1]{#2}\endgroup}

\begin{document}

\bibliographystyle{hieeetr}

\begin{titlepage}
\begin{flushright}
CP3-06-02
\end{flushright}

\vskip 2.cm

\begin{center}
{\Large\bf Color-dressed recursive relations \\
\bigskip for multi-parton amplitudes}\\
\medskip
\bigskip\bigskip\bigskip\bigskip
{\large  Claude Duhr, Stefan H{\"o}che\symbolfootnote[1]{
  permanent address: \parbox[t]{0.5\textwidth}{
    Institut f{\"u}r theoretische Physik\\
    Technische Universit{\"a}t Dresden\\
    D-01062 Dresden, Germany}}, Fabio Maltoni}\\
\medskip\bigskip\bigskip

{Institut de Physique Th\'{e}orique and \\
Centre for Particle Physics and Phenomenology (CP3) \\
Universit\'{e} Catholique de Louvain\\[1mm]
Chemin du Cyclotron 2 \\
B-1348 Louvain-la-Neuve, Belgium
}\\
\end{center}

\bigskip
\begin{center}
\today
\end{center}

\bigskip\bigskip\bigskip

\begin{abstract}\noindent
Remarkable progress inspired by twistors has lead to very simple
analytic expressions and to new recursive relations for 
multi-parton color-ordered amplitudes. We show how such 
relations can be extended to include color and
present the corresponding color-dressed formulation for the
Berends-Giele, BCF and a new kind of CSW recursive relations. 
A detailed comparison of the numerical efficiency of the different
approaches to the calculation of multi-parton cross sections 
is performed.
\end{abstract}
\end{titlepage}

\baselineskip 16pt

\section{Introduction}
\label{sec:Intro}

The quest for physics beyond the standard model will soon enter an
exciting phase as the LHC starts colliding protons at 14 TeV. At such
an energy we will be probing the mechanism which breaks the electroweak
symmetry and hopefully understand what stabilizes the associated
scale. Whether this mechanism can be described perturbatively by a
standard model Higgs or can be related to new symmetries (such as
supersymmetry) or even to the discovery of an extended space-time
structure, is the subject of an intense theoretical activity. The
possibility, however, of answering such far reaching questions at the
LHC will eventually depend on our ability to discriminate 
signals of new physics from large standard model backgrounds.  With so
much energy available in the center of mass, the cross section for multi-jet QCD events even
associated with heavy objects such as weak bosons or top quarks will be in
most cases very large if not overwhelming. 

As the need for better predictions for QCD processes with many particles in
the final state has become clear, a substantial activity on developing
the techniques and the tools to perform such calculations has spurred.
Several codes have by now become available that can compute, 
{\it numerically}, tree-level cross sections and generate events 
with many particles in the final state in an 
automatic way~\cite{Mangano:2002ea,Maltoni:2002qb,Gleisberg:2003xi,Papadopoulos:2005ky}. Progress has been significant also in improving the accuracy 
of the predictions by calculating next-to- and next-to-next-to-leading 
QCD corrections for processes with up to three and just one particle 
in the final state respectively~\cite{Dixon:2005sp}.

In the midst of this effort, unexpected theoretical progress has come
from the so-called twistor-inspired methods~\cite{Witten:2003nn}, which have
provided new techniques to compute analytic results for gauge
amplitudes both at tree and one-loop level~\cite{Cachazo:2005ga}. 
These new methods go back to a correspondence between a weakly coupled 
$\begin{cal}N\end{cal}=4$ Super Yang-Mills theory and a certain type of string theory. 
The key point in this correspondence is that all tree-level (color-ordered) 
amplitudes are related to algebraic curves in twistor space. In 
Ref.~\cite{Cachazo:2004kj} it was shown that this leads to the so-called 
CSW rules, which state that all tree-level (color-ordered) amplitudes can 
be constructed using a small class of very special amplitudes, the 
maximally helicity violating (MHV) amplitudes. Another 
important result has been the formulation of a new kind of recursive
relations in addition to the well-known Berends-Giele 
recursion~\cite{Berends:1987me}, the so-called BCF 
relations~\cite{Britto:2004ap,Britto:2005fq}.
They state that any tree-level (color-ordered) 
amplitude can be constructed from products of two on-shell amplitudes 
of fewer particles, multiplied by a simple scalar propagator. 
These new calculational tools have allowed to derive  
expressions for some multi-parton 
amplitudes~\cite{Cachazo:2004kj, Luo:2005my, deFlorian:2006ek} 
and certain classes of splitting functions~\cite{Birthwright:2005ak, Birth:2005vi}, which have simple and compact analytic forms.
So far, many extensions of the twistor-inspired methods have been presented, 
in particular generalizations to include scalars~\cite{Dixon:2004za}, 
fermions~\cite{Luo:2005rx, Badger:2005zh} and photons~\cite{Ozeren:2005mp}.

The purpose of
this work is to present a new extension of the twistor-inspired methods.
We show that it is possible to reformulate the
twistor-inspired recursive relations for the color-ordered amplitudes
in terms of the full amplitudes. The motivations are twofold.  The
first is mostly theoretical and stems from the observation of an
interesting similarity between a color decomposition based on the
adjoint representation and the BCF recursive relations which suggested
the existence of a formulation that embodies both. The second is more
pragmatic and aims at establishing whether the new twistor-inspired
recursive relations are an improvement also at the numerical level.
In order to make a consistent comparison with the most efficient
algorithms available~\cite{Mangano:2002ea,Draggiotis:2002hm}, a
recursive formulation which includes color is necessary.  In the
standard approach where the color-ordered amplitudes are calculated
analytically (or numerically), one has to sum over the permutations of
the color orderings to obtain the full amplitude.  This algorithm has
an intrinsic factorial growth and cannot compete with the available
numerical methods which only grow exponentially.  In this work we
derive a general method to reinstate color into the recursive
relations for color-ordered amplitudes and apply it to the
BCF and to a modified version of the CSW
relations.

The paper is organized as follows.  In \Sec{sec:coldec} we review the
notion of color decompositions, highlighting the different features of
the various color bases available.  In \Sec{sec:BG} we
discuss the Berends-Giele recursive relations and we present a simple
derivation of their color-dressed counterpart which serves as an
illustration of the method that will be applied later. In
Section~\ref{sec:CDBCF} we prove our first main result, {\it i.e.}, the
color-dressed version of the BCF relations, Eq.~(\ref{eq:CDBCF})  and
discuss its most important features.  In Section~\ref{sec:CDCSW}  we
reformulate the CSW relations in terms of simple new three-point
effective vertices and derive their color-dressed version.
Section~\ref{sec:results} contains the numerical results on the
evaluation of multi-gluon amplitudes obtained using the different
color-dressed recursive relations, focusing on the comparison with 
known techniques. Finally we draw our conclusions.

\section{Color Decompositions}
\label{sec:coldec}

In this section we briefly review the notion of color decomposition of tree-level QCD
amplitudes and the available results. Emphasis is given to those aspects that will
play an important role in the following.

The basic idea of a color decomposition is to factorize the information
on the gauge structure from the kinematics. As an example, consider the
amplitude for $n$ gluons of colors $a_1, a_2, \ldots, a_n$ ($a_i = 1, \ldots,
N^2 - 1$).  One can easily prove that at tree level, such an amplitude can 
be decomposed as~\cite{Mangano:1987xk}
\begin{equation}
\label{eq:fund_decomposition} {\cal A}(1,\ldots,n) = \sum_{\sigma \in S_{n-1}}
{\rm Tr}\,
(T^{a_1} T^{a_{\sigma_2}} \cdots T^{a_{\sigma_n}})\; A(1,\sigma_2,\ldots,\sigma_n)\;,
\end{equation}
where $T^{a}$ are the fundamental-representation matrices of $SU(N)$,
and the sum is over all $(n-1)!$ permutations of $(2,\ldots,n)$.  Each trace
corresponds to a particular color structure.  The factor associated with each
color structure, $A$, is called a color-ordered amplitude.\footnote{Also referred to
as a dual amplitude or partial amplitude.} It depends on the
four-momenta $p_i$ and polarization vectors $\epsilon_i$ of the $n$ gluons,
represented simply by $i$ in its argument. The
color-ordered amplitudes are far simpler to calculate than the full amplitude ${\cal A}$ 
due to the smaller number of Feynman diagrams contributing to them.
They have several remarkable properties. Among them, a 
special role is played by the so-called Kleiss-Kuijf relations~\cite{Kleiss:1988ne}.
These are linear relations amongst the amplitudes directly inherited from
the gauge structure,{\it i.e.,} from color, which in the case of $n$-gluon amplitudes reduce
the number of linearly-independent amplitudes to $(n-2)!$. It is then clear that the number 
of terms in Eq.~(\ref{eq:fund_decomposition}) is not minimal.
This decomposition has no special feature except that it was inspired by string
theories (color factors are the Chan-Paton factors of the open strings). It is 
universally used to illustrate the idea of color decomposition and to define
the color-ordered amplitudes $A$ in terms of the full amplitude ${\cal A}$.
It can be shown, however, that this definition does not depend on the color basis.

Recently, another decomposition has been introduced, which is based
on color flows~\cite{Kanaki:2000ey,Maltoni:2002mq}. 
This decomposition arises when treating the $SU(N)$ gluon field
as an $N \times N$ matrix $(A_{\mu})^{i}_{j}$ ($i,j = 1,\ldots, N$), rather
than as a one-index field $A_{\mu}^{a}$ ($a = 1,\ldots, N^2-1$).  The $n$-gluon
amplitude may then be decomposed as
\begin{equation}
\label{eq:colorflow_decomposition} {\cal A}(1,\ldots,n) = \sum_{\sigma \in S_{n-1}}
\delta_{i_1}^{\bar\jmath_{\sigma_2}} \delta_{i_{\sigma_2}}^{\bar \jmath_{\sigma_3}} \cdots 
\delta_{i_{\sigma_{n}}}^{\bar \jmath_1} \; A(1,\sigma_2,\ldots,\sigma_n)\;,
\end{equation}
where the sum is over all $(n-1)!$ permutations of $(2,\ldots,n)$.  The
partial amplitudes that appear in this decomposition are the same as in the
decomposition in the fundamental representation. The color-flow decomposition has several nice features,
the most attractive one being its simplicity. In fact, 
it is a very natural way to decompose a QCD amplitude, and as
the name suggests, it is based on the flow of color, so the decomposition
has a simple physical interpretation. It does not involve any color matrix
and the color factors in front of each amplitude are either zero or one. 
This basis makes the numerical calculation of color factors in the evaluation of full
amplitudes fast, even though the number of terms in the 
sum over the $(n-1)!$ permutations is not minimal. Finally,
such a decomposition exists for all tree-level parton amplitudes 
including any number of quark pairs and gluons.

A third decomposition of the multi-gluon amplitude is available, which is
based on the adjoint representation of $SU(N)$ rather than the
fundamental representation \cite{DelDuca:1999ha,DelDuca:1999rs}.  
The $n$-gluon amplitude in this decomposition may be written as
\begin{equation}
\label{eq:adjoint_decomposition} {\cal A}(1,\ldots,n) = \sum_{\sigma \in S_{n-2}}
(F^{a_{\sigma_2}} F^{a_{\sigma_3}} \cdots 
 F^{a_{\sigma_{n-1}}})_{a_1 a_n} \; A(1,\sigma_2,\ldots,\sigma_{n-1},n)\;,
\end{equation}
where $(F^a)^b_c = -if^{a b c}$ are the adjoint-representation matrices of
$SU(N)$ ($f^{abc}$ are the structure constants), and the sum is over all
$(n-2)!$ permutations of $(2,\ldots,n-1)$. The indices corresponding to 
the first and the last gluon are taken as ``references'' and are not 
included in the permutations. 
The partial amplitudes that appear in this decomposition are the 
same as in the other decomposition, but only the $(n-2)!$ linearly-independent 
amplitudes are needed. In this respect this formulation is ``minimal'' as
there is no redundancy and the Kleiss-Kuijf relations are embodied in the color factors.
As we will elaborate upon in the following, there exists a remarkable formal similarity with
the BCF recursive relations, where two gluons are also taken as a reference to build
up the full amplitude.

\section{Color-dressed Berends-Giele relations}
\label{sec:BG}

In Ref.~\cite{Berends:1987me}, Berends and Giele introduced the color-ordered 
$n$-point gluon off-shell current $ J^\mu$, which can be defined as the 
sum of all color-ordered Feynman diagrams with $n$ external on-shell legs 
and a single off-shell leg with polarization $\mu$. 
The color-ordered off-shell currents can be constructed using the 
Berends-Giele recursive relations
\begin{eqnarray}\label{eq:BG}
 J^\mu(1,2,\ldots,n)=  \frac{-i}{P_{1,n}^2}& \Bigg\{ &\sum_{k=1}^{n-1}V_3^{\mu\nu\rho}\left(P_{1,k},P_{k+1,n}\right)J_\nu(1,\ldots,k)J_\rho(k+1,\ldots,n)\\
& + & \sum_{j=1}^{n-2}\sum_{k=j+1}^{n-1}V_4^{\mu\nu\rho\sigma}J_\nu(1,\ldots,j)J_\rho(j+1,\ldots,k)J_\sigma(k+1,\ldots,n)\Bigg\},\nonumber
\end{eqnarray} 
where 
\begin{equation}
P_{i,j}=p_i+p_{i+1}+\ldots+p_{j-1}+p_j,
\end{equation}
and $V_3^{\mu\nu\rho}\left(P_{1,k},P_{k+1,n}\right)$ and 
$V_4^{\mu\nu\rho\sigma}$ are the color-ordered three and four-gluon vertices defined in Ref.~\cite{Dixon:1996wi}. 
It is easy to see that the four-gluon vertex appearing in these relations 
introduces a larger number of possible combinations of subcurrents than
the three-gluon vertex. It is possible to simplify the recursion 
by decomposing all four-gluon vertices into three-vertices including a 
tensor particle (Fig.~\ref{fig:tensor4gluonCO}). Using this decomposition, 
the Berends-Giele recursive relations can be rewritten such that only 
three-point vertices are present
\begin{eqnarray}\label{eq:BGtensor}
&&\text{\hspace*{-5ex}}J^\mu(1,2,\ldots,n)=  \frac{-i}{P_{1,n}^2}  \sum_{k=1}^{n-1}\Bigg\{V_3^{\mu\nu\rho}\left(P_{1,k},P_{k+1,n}\right)J_\nu(1,\ldots,k)J_\rho(k+1,\ldots,n)\\
&&\quad +\; V_T^{\nu\mu\alpha\beta}J_\nu(1,\ldots,k)J_{\alpha\beta}(k+1,\ldots,n)+ V_T^{\mu\sigma\alpha\beta}J_{\alpha\beta}(1,\ldots,k)J_{\sigma}(k+1,\ldots,n)\Bigg\},\nonumber
\end{eqnarray} 
where $J_{\alpha\beta}$ is a tensor off-shell current, and 
$V_T^{\mu\nu\alpha\beta}$ is the tensor-gluon vertex, defined as
\begin{equation}
V_T^{\mu\nu\rho\sigma}=\frac{ig}{2}\left(g^{\mu\rho}g^{\nu\sigma}-g^{\mu\sigma}g^{\nu\rho}\right).
\end{equation}
As there exists no one-point tensor off-shell current, 
all such currents appearing in Eq.~(\ref{eq:BGtensor}) are defined as zero. 
The tensor off-shell currents can be easily constructed recursively from gluon 
off-shell currents
\begin{equation}
J_{\mu\nu}(1,2,\ldots,n)=iD_{\mu\nu\alpha\beta}\ V_T^{\sigma\rho\alpha\beta}\ \sum_{k=1}^{n-1}J_\rho(1,\ldots,k)J_\sigma(k+1,\ldots,n),
\label{eq:tensorRR}
\end{equation}
where $iD_{\mu\nu\alpha\beta}$ is the color-ordered tensor ``propagator'', 
defined as
\begin{equation}
iD_{\mu\nu\rho\sigma}=-\frac{i}{2}\left(g^{\mu\rho}g^{\nu\sigma}-g^{\mu\sigma}g^{\nu\rho}\right).
\end{equation}
\myfigure{t!}{
\fcolorbox{white}{white}{
  \begin{picture}(255,60) (76,-30)
    \SetWidth{0.5}
    \SetColor{Black}
    \Line(165,30)(180,0)
    \Line(180,0)(165,-30)
    \Line(225,30)(210,0)
    \Line(210,0)(225,-30)
    \DashLine(180,0)(210,0){6}
    \Line(271,30)(301,15)
    \Line(301,14)(331,30)
    \Line(271,-30)(301,-15)
    \Line(301,-15)(331,-30)
    \DashLine(301,15)(301,-15){6}
    \Text(244,-2)[lb]{\Large{\Black{$+$}}}
    \Text(137,-1)[lb]{\Large{\Black{$=$}}}
    \Line(76,30)(136,-30)
    \Line(76,-30)(136,30)
  \end{picture}
}}{
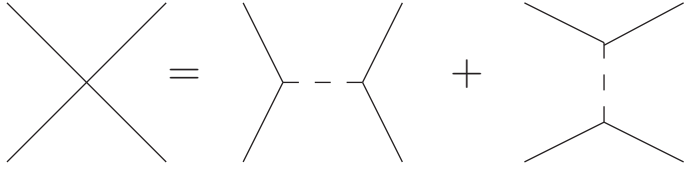
Diagrammatic representation of the decomposition of the color-ordered four-gluon vertex.\label{fig:tensor4gluonCO}}

We now present a systematic method to dress color-ordered 
recursive relations with color in order to obtain recursive relations for 
the color-dressed off-shell currents. In the color-flow decomposition, 
a color-dressed gluon off-shell current can be written as
\begin{equation}
\begin{cal}J\end{cal}^\mu_{I\bar J}(1,2,\ldots,n)=\sum_{\sigma \in S_n}\delta_{i_{\sigma_1}}^{\bar{J}}\delta_{i_{\sigma_2}}^{\bar{\jmath}_{\sigma_1}}\ldots\delta_{I}^{\bar{\jmath}_{\sigma_n}}J^\mu(\sigma_1,\sigma_2,\ldots,\sigma_n),
\label{eq:colorFlowOffShell}
\end{equation}
where $(I,\bar{J})$ is the color of the off-shell leg. A color-dressed tensor off-shell current can be obtained similarly. 
We will explain the color dressing of the Berends-Giele recursive 
relations, Eq.~(\ref{eq:BGtensor}), dealing with 
the pure gluon vertices and the tensor-gluon vertices separately. 
After inserting Eq.~(\ref{eq:BGtensor}), into the color-flow decomposition, 
Eq.~(\ref{eq:colorFlowOffShell}), the three-gluon vertex part reads
\begin{equation}
\frac{-i}{P_{1,n}^2}\sum_{\sigma \in S_{n}}\ \sum_{k=1}^{n-1}\delta_{i_{\sigma_1}}^{\bar{J}}\delta_{i_{\sigma_2}}^{\bar{\jmath}_{\sigma_1}}\ldots\delta_{I}^{\bar{\jmath}_{\sigma_n}}V^{\mu\nu\rho}_3\left(P_{\sigma_1,\sigma_k},P_{\sigma_{k+1},\sigma_n}\right) J_\nu(\sigma_1,\ldots,\sigma_k)J_\rho(\sigma_{k+1},\ldots,\sigma_{n}),
\label{eq:BG3V}
\end{equation}
where 
\begin{equation}
\begin{split}
P_{\sigma_1,\sigma_k}&= p_{\sigma_1}+p_{\sigma_2}+\ldots+p_{\sigma_k},\\
P_{\sigma_{k+1},\sigma_n}&= p_{\sigma_{k+1}}+p_{\sigma_{k+2}}+\ldots+p_{\sigma_n}.
\end{split}
\end{equation}
The color factor appearing in Eq.~(\ref{eq:BG3V}) can be written as (Fig.~\ref{fig:BG3V})
\begin{equation}
\delta_{i_{\sigma_1}}^{\bar{J}}\delta_{i_{\sigma_2}}^{\bar{\jmath}_{\sigma_1}}\ldots\delta_{I}^{\bar{\jmath}_{\sigma_n}}=
 \delta_{G}^{\bar{J}}\delta_{I}^{\bar{H}}\quad \delta_{L}^{\bar{G}}\delta_{N}^{\bar{K}}\delta_{H}^{\bar{M}}\quad \delta_{i_{\sigma_1}}^{\bar{L}}\ldots\delta_{K}^{\bar{\jmath}_{\sigma_k}}\quad \delta_{i_{\sigma_{k+1}}}^{\bar{N}}\ldots\delta_{M}^{\bar{\jmath}_{\sigma_n}},
\label{eq:ColorFactDecomp3V}
\end{equation}
where
\begin{itemize}
\item[-] $\delta_{G}^{\bar{J}}\delta_{I}^{\bar{H}}$ is the color structure of the propagator appearing in the Berends-Giele recursive relations. 
\item[-] $\delta_{i_{\sigma_1}}^{\bar{L}}\ldots\delta_{K}^{\bar{\jmath}_{\sigma_k}}$ is the color structure of the subcurrent $J_\nu(\sigma_1,\ldots,\sigma_k)$, where the off-shell leg $\nu$ has color $(K,\bar L)$.
\item[-] $\delta_{i_{\sigma_{k+1}}}^{\bar{N}}\ldots\delta_{M}^{\bar{\jmath}_{\sigma_n}}$ is the color structure of the subcurrent $J_\rho(\sigma_{k+1},\ldots,\sigma_{n})$, where the off-shell leg $\rho$ has color $(M,\bar N)$.
\item[-] $\delta_{L}^{\bar{G}}\delta_{N}^{\bar{K}}\delta_{H}^{\bar{M}}$ is part of the color structure of a three-gluon vertex to which the off-shell legs $\mu$, $\nu$, $\rho$ with colors $(G,\bar H)$, $(K,\bar L)$, $(M,\bar N)$ are attached.
\end{itemize}
\myfigure{t!}{
\fcolorbox{white}{white}{
  \begin{picture}(356,201) (70,-130)
    \SetWidth{0.8}
    \SetColor{Black}
    \LongArrow(161,-31)(208,-31)
    \SetWidth{0.5}
    \GOval(127,-32)(20,22)(0){0.882}
    \Text(70,-28)[lb]{\normalsize{\Black{$\bar J$}}}
    \Text(70,-44)[lb]{\normalsize{\Black{$I$}}}
    \ArrowLine(80,-39)(106,-39)
    \ArrowLine(106,-24)(80,-24)
    \Text(117,-39)[lb]{\Large{\Black{$\mathcal{J}^\mu$}}}
    \ArrowLine(392,-92)(376,-77)
    \ArrowLine(365,-89)(381,-104)
    \GOval(402,-110)(20,22)(0){0.882}
    \Text(357,-88)[lb]{\normalsize{\Black{$M$}}}
    \Text(372,-75)[lb]{\normalsize{\Black{$\bar{N}$}}}
    \Text(396,-116)[lb]{\Large{\Black{$\mathcal{J}^\rho$}}}
    \Text(349,8)[lb]{\normalsize{\Black{$L$}}}
    \Text(362,-4)[lb]{\normalsize{\Black{$\bar{K}$}}}
    \Text(263,-44)[lb]{\normalsize{\Black{$H$}}}
    \Text(263,-29)[lb]{\normalsize{\Black{$\bar{G}$}}}
    \Text(360,-67)[lb]{\normalsize{\Black{$N$}}}
    \Text(347,-78)[lb]{\normalsize{\Black{$\bar{M}$}}}
    \Text(216,-28)[lb]{\normalsize{\Black{$\bar J$}}}
    \ArrowLine(226,-39)(243,-39)
    \ArrowLine(243,-24)(226,-24)
    \Text(216,-44)[lb]{\normalsize{\Black{$I$}}}
    \Text(247,-28)[lb]{\normalsize{\Black{$G$}}}
    \Text(247,-44)[lb]{\normalsize{\Black{$\bar{H}$}}}
    \ArrowLine(378,16)(394,31)
    \ArrowLine(380,42)(365,26)
    \GOval(402,51)(20,22)(0){0.882}
    \Text(370,9)[lb]{\normalsize{\Black{$K$}}}
    \Text(355,21)[lb]{\normalsize{\Black{$\bar{L}$}}}
    \Text(395,45)[lb]{\Large{\Black{$\mathcal{J}^\nu$}}}
    \ArrowLine(316,-24)(271,-24)
    \ArrowLine(271,-39)(316,-39)
    \ArrowLine(316,-39)(346,-69)
    \ArrowLine(346,6)(316,-24)
    \ArrowLine(358,-58)(331,-31)
    \ArrowLine(331,-31)(358,-4)
  \end{picture}
}}{
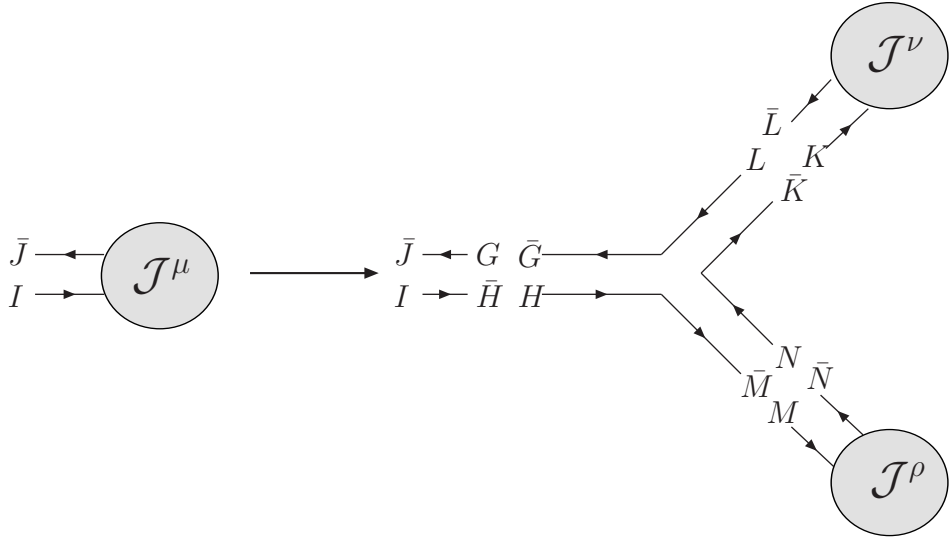
Diagrammatic representation of the decomposition~(\ref{eq:ColorFactDecomp3V}) of the color factor of the three-gluon vertex part.\label{fig:BG3V}}
 
We now define an ordered partition
of a set $E$ into two independent parts as a pair $(\pi_1,\pi_2)$ 
of subsets of $E$ such that $\pi_1\oplus\pi_2=E$, which means $(\pi_1,\pi_2)\neq (\pi_2,\pi_1)$.  
Furthermore, we call (unordered) partition of a set $E$ into two 
independent parts a set $\{\pi_1,\pi_2\}$ of subsets of $E$ such that 
$\pi_1\oplus\pi_2=E$ and $\{\pi_1,\pi_2\}=\{\pi_2,\pi_1\}$ . 
These definitions can be easily 
extended to partitions of a set $E$ into $n>2$ independent parts, for 
both the ordered and the unordered case.

In the case encountered here $E=\{1,2,\ldots,n\}$. We will denote 
the set of all ordered partitions of $E$ into two independent parts 
by $OP(n,2)$ and the set of all (unordered) partitions of $E$ 
into two independent parts by $P(n,2)$.
Using these definitions, the sum over permutations appearing in 
Eq.~(\ref{eq:BG3V}) can be decomposed as follows: For a given value of $k$,
\begin{itemize}
\item[-] Choose an ordered partition $\pi=(\pi_1,\pi_2)$ in $OP(n,2)$ such that $\#\pi_1=k$, where $\#\pi_1$ is the number of elements in the set $\pi_1$.
\item[-] Fix the first $k$ elements of the permutation to be in the subset $\pi_1$.
\item[-] Sum over all permutations of the first $k$ elements and over all permutations of the last $n-k$ elements.
\item[-] Sum over all possible choices for the ordered partition $\pi=(\pi_1,\pi_2)$.
\end{itemize}
This is equivalent to the replacement
\begin{equation}
\sum_{k=1}^{n-1}\ \sum_{\sigma \in S_{n}}\rightarrow \sum_{\pi \in OP(n,2)}\ \sum_{\sigma \in S_k} \ \sum_{\sigma ' \in S_{n-k}}.
\label{eq:OP3V}
\end{equation}
The three-gluon vertex part now reads
\begin{eqnarray}\delta_{G}^{\bar{J}}\delta_{I}^{\bar{H}}\ \frac{-i}{P_{1,n}^2}\sum_{\pi \in OP(n,2)}\ \sum_{\sigma \in S_{k}}\sum_{\sigma '\in S_{n-k}}\ \delta_{L}^{\bar{G}}\delta_{N}^{\bar{K}}\delta_{H}^{\bar{M}}\ V^{\mu\nu\rho}_3\left(P_{\sigma_{\pi^1},\sigma_{\pi^k}},P_{\sigma '_{\pi^{k+1}},\sigma '_{\pi^n}}\right)\nonumber \\
\delta_{i_{\sigma_{\pi^1}}}^{\bar{L}}\ldots\delta_{K}^{\bar{\jmath}_{\sigma_{\pi^k}}}J_\nu(\sigma_{\pi^1},\ldots,\sigma_{\pi^k})\ \delta_{i_{\sigma '_{\pi^{k+1}}}}^{\bar{N}}\ldots\delta_{M}^{\bar{\jmath}_{\sigma '_{\pi^n}}}J_\rho(\sigma '_{\pi^{k+1}},\ldots,\sigma '_{\pi^n}),
\end{eqnarray}
where $\pi_1=\{\pi^1,\pi^2\ldots,\pi^k\}$ and $\pi_2=\{\pi^{k+1},\pi^{k+2},\ldots,\pi^{n}\}$.\\
Clearly, $P_{\sigma_{\pi^1},\sigma_{\pi^k}}$ and 
$P_{\sigma '_{\pi^{k+1}},\sigma '_{\pi^n}}$ only depend on the choice 
of the ordered partition $\pi=(\pi_1,\pi_2)$, but not on the order of 
the elements in $\pi_1$ and $\pi_2$. We therefore define
\begin{equation}
\begin{split}
P_{\pi_1} &= p_{\pi^1}+p_{\pi^2}+\ldots+p_{\pi^k},\\
P_{\pi_2} &= p_{\pi^{k+1}}+p_{\pi^{k+2}}+\ldots+p_{\pi^n}.
\end{split}
\end{equation}
It is now possible to identify several subcurrents in this expression, namely
\begin{eqnarray}
\begin{cal}J\end{cal}^{K\bar{L}}_\nu(\pi_1) & = & \sum_{\sigma \in S_k}\delta_{i_{\sigma_{\pi^1}}}^{\bar{L}}\ldots\delta_{K}^{\bar{\jmath}_{\sigma_{\pi^k}}}J_\nu(\sigma_{\pi^1},\ldots,\sigma_{\pi^k}),\\
\begin{cal}J\end{cal}^{M\bar N}_\rho(\pi_2) & = & \sum_{\sigma ' \in S_{n-k}}\ \delta_{i_{\sigma '_{{\pi^{k+1}}}}}^{\bar{N}}\ldots\delta_{M}^{\bar{\jmath}_{\sigma '_{\pi^n}}}J_\rho(\sigma '_{\pi^{k+1}},\ldots,\sigma '_{\pi^n}),
\end{eqnarray}
so that the three-gluon vertex part reads
\begin{equation}
\delta_{G}^{\bar{J}}\delta_{I}^{\bar{H}}\ \frac{-i}{P_{1,n}^2} \sum_{\pi \in OP(n,2)} \delta_{L}^{\bar{G}}\delta_{N}^{\bar{K}}\delta_{H}^{\bar{M}}\ V^{\mu\nu\rho}_3\left(P_{\pi_1},P_{\pi_2}\right)\ \begin{cal}J\end{cal}^{K\bar L}_\nu(\pi_1)\ \begin{cal}J\end{cal}^{M\bar N}_\rho(\pi_2).
\end{equation}
In Ref.~\cite{Maltoni:2002mq} it was shown that the (color-dressed) three-gluon vertex can be expressed in the color-flow decomposition as\footnote{For brevity, the color indices of the vertex are not written explicitly.}
\begin{equation}
\begin{cal}V\end{cal}_3^{\mu\nu\rho}\left(P_{\pi_1},P_{\pi_2}\right)=\delta_{L}^{\bar{G}}\delta_{N}^{\bar{K}}\delta_{H}^{\bar{M}}\ V^{\mu\nu\rho}_3\left(P_{\pi_1},P_{\pi_2}\right)+\delta_{N}^{\bar{G}}\delta_{L}^{\bar{M}}\delta_{H}^{\bar{K}}\ V^{\mu\rho\nu}_3\left(P_{\pi_2},P_{\pi_1}\right).
\label{eq:CD3vertex}
\end{equation}
So, finally the three-gluon vertex part can be written as
\begin{equation}
\delta_{G}^{\bar{J}}\delta_{I}^{\bar{H}}\ \frac{-i}{P_{1,n}^2} \sum_{\pi \in P(n,2)}\begin{cal}V\end{cal}_3^{\mu\nu\rho}\left(P_{\pi_1},P_{\pi_2}\right)\begin{cal}J\end{cal}^{K\bar L}_\nu(\pi_1)\ \begin{cal}J\end{cal}^{M\bar N}_\rho(\pi_2).
\end{equation}

We now address the tensor-gluon part in the Berends-Giele recursive 
relations, Eq.~(\ref{eq:BGtensor}). From Fig.~\ref{fig:4GVcolorstruc} 
it can be seen that for the $s$-channel appearing in the decomposition 
of the four-gluon vertex, each tensor-gluon vertex has the same color-flow 
structure as a three-gluon vertex. The $t$-channel contribution is similar. 
Therefore the tensor-gluon vertex can be written in the color-flow decomposition as
\myfigure{t!}{
\fcolorbox{white}{white}{
  \begin{picture}(337,84) (30,-90)
    \SetWidth{0.5}
    \SetColor{Black}
    \ArrowLine(62,-48)(30,-16)
    \ArrowLine(30,-80)(61,-48)
    \ArrowLine(40,-6)(72,-38)
    \ArrowLine(71,-58)(40,-90)
    \ArrowLine(72,-37)(104,-6)
    \ArrowLine(113,-16)(82,-48)
    \ArrowLine(82,-48)(114,-80)
    \ArrowLine(103,-89)(72,-58)
    \SetWidth{0.7}
    \LongArrow(114,-48)(156,-48)
    \SetWidth{0.5}
    \ArrowLine(251,-37)(283,-37)
    \ArrowLine(283,-58)(251,-58)
    \ArrowLine(199,-48)(167,-16)
    \ArrowLine(167,-80)(198,-48)
    \ArrowLine(178,-6)(210,-38)
    \ArrowLine(209,-58)(178,-90)
    \ArrowLine(208,-37)(240,-37)
    \ArrowLine(240,-58)(208,-58)
    \ArrowLine(292,-37)(324,-37)
    \ArrowLine(324,-58)(292,-58)
    \ArrowLine(325,-37)(357,-6)
    \ArrowLine(366,-16)(335,-48)
    \ArrowLine(335,-48)(367,-80)
    \ArrowLine(356,-89)(325,-58)
  \end{picture}
}}{
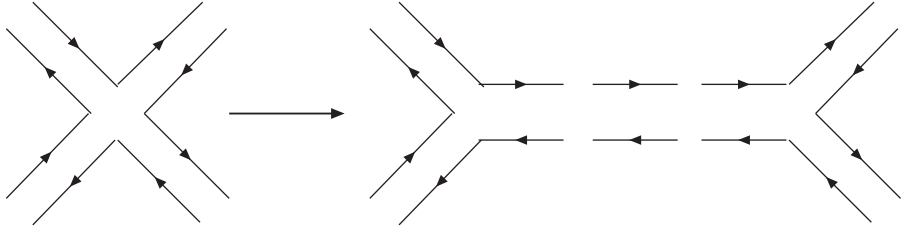
Color-flow structure of the $s$-channel contribution to the for gluon vertex.
\label{fig:4GVcolorstruc}}
\begin{equation}
\begin{cal}V\end{cal}_T^{\mu\nu\alpha\beta}=\delta_{i_1}^{\bar J}\delta_{i_2}^{\bar{\jmath}_1}\delta_{I}^{\bar{\jmath}_2}\ V_T^{\mu\nu\alpha\beta}
+\delta_{i_2}^{\bar J}\delta_{i_1}^{\bar{\jmath}_2}\delta_{I}^{\bar{\jmath}_1}\ V_T^{\nu\mu\alpha\beta},
\label{eq:VTColorFlow}
\end{equation}
where $(I,\bar{J})$ is the color of the tensor particle.
The color dressing of the tensor-gluon part in Eq.~(\ref{eq:BGtensor})
is hence exactly the same as for the pure gluon part, leading to 
\begin{equation}
\sum_{\pi \in OP(n,2)}\delta_{L}^{\bar G}\delta_{N}^{\bar K}\delta_{H}^{\bar M}\ \left\{\;V_T^{\nu\mu\alpha\beta}\begin{cal}J\end{cal}_\nu^{K\bar L}(\pi_1)\begin{cal}J\end{cal}_{\alpha\beta}^{M\bar N}(\pi_2)+V_T^{\mu\nu\alpha\beta}\begin{cal}J\end{cal}_{\alpha\beta}^{K\bar L}(\pi_1)\begin{cal}J\end{cal}_{\nu}^{M\bar N}(\pi_2)\;\right\}.
\end{equation}
As the sum runs over all elements in $OP(n,2)$, we can exchange 
$\pi_1$ and $\pi_2$ as well as the color indices 
$(K,\bar L)$ and $(M,\bar N)$ in the last term. 
Using Eq.~(\ref{eq:VTColorFlow}) the tensor part now becomes
\begin{equation}
\sum_{\pi \in OP(n,2)}\begin{cal}V\end{cal}_T^{\mu\nu\alpha\beta}\begin{cal}J\end{cal}_\nu^{K\bar L}(\pi_1)\begin{cal}J\end{cal}_{\alpha\beta}^{M\bar N}(\pi_2).
\end{equation}
Hence the color-dressed recursive relations with all four-gluon 
vertices replaced by tensor particles read
\begin{eqnarray}
\begin{cal}J\end{cal}^{I\bar J}_\mu(1,\ldots,n)=D_{\mu\nu}\left(P_{1,n}\right)&\Bigg[&\sum_{\pi \in P(n,2)}\begin{cal}V\end{cal}_3^{\nu\rho\sigma}\left(P_{\pi_1},P_{\pi_2}\right)\ \begin{cal}J\end{cal}_\rho^{K\bar L}(\pi_1)\begin{cal}J\end{cal}_\sigma^{M\bar N}(\pi_2)\nonumber \\
&+&\sum_{\pi \in OP(n,2)}\begin{cal}V\end{cal}_T^{\mu\rho\alpha\beta}\,\begin{cal}J\end{cal}_\rho^{K\bar L}(\pi_1)\,\begin{cal}J\end{cal}_{\alpha\beta}^{M\bar N}(\pi_2)\Bigg].
\label{eq:CDBGT}
\end{eqnarray}

To complete the color dressing of the Berends-Giele recursive relations, we have to apply the color-dressing method introduced above to the recursive relations for the off-shell tensor-currents, Eq.~(\ref{eq:tensorRR}). Both the color-dressed vertex and the color-dressed off-shell tensor-current have the same form as in the pure gluon case and the recursive relations for the tensor particle, Eq.~(\ref{eq:tensorRR}), have the same structure as for the three-gluon vertex part in the previous section. Therefore, one can immediately write down the color-dressed recursive relations for the off-shell tensor-current
\begin{equation}
\begin{cal}J\end{cal}_{\mu\nu}^{I\bar J}(1,2,\ldots,n)=iD_{\mu\nu\alpha\beta}\ \begin{cal}V\end{cal}_T^{\sigma\rho\alpha\beta}\ \sum_{\pi \in P(n,2)}\begin{cal}J\end{cal}_\rho^{K\bar L}(\pi_1)\begin{cal}J\end{cal}_\sigma^{M\bar N}(\pi_2).
\label{eq:CDtensorRR}
\end{equation}
The two recursive relations, Eq.~(\ref{eq:CDBGT}) and Eq.~(\ref{eq:CDtensorRR}), can be solved simultaneously to construct color-dressed gluon off-shell currents for arbitrary $n$. The full color-dressed scattering amplitude is then recovered be putting the off-shell leg on-shell. This result is equivalent to the Dyson-Schwinger algorithm presented in Ref.~\cite{Papadopoulos:2005ky}. It should be noticed that the color-dressed recursive relations have the same form as the color-ordered Berends-Giele recursive relations, Eq.~(\ref{eq:BGtensor}) and Eq.~(\ref{eq:tensorRR}). The only difference between the color-ordered and the color-dressed case is that in the latter we sum over unordered objects and no permutations need to be taken into account. This is a general feature which will turn out to be common to all color-dressed recursive relations.

\section{Color-dressed BCF relations}
\label{sec:CDBCF}

In this section, we apply the same method employed to construct the 
color-dressed Berends-Giele recursive relations to the BCF recursive 
relations, presented in Ref.~\cite{Britto:2004ap, Britto:2005fq}. 
Assuming that gluons $1$ and $n$ have opposite helicities, the BCF 
recursive relations read
\begin{equation}
A_n(1,2,\ldots,n)=\sum_{k=2}^{n-2}A_{k+1}\left(\hat 1,2,\ldots,k,-\hat{P}_{1,k}^{-h}\right)\frac{1}{P_{1,k}^2}A_{n-k+1}\left(\hat{P}_{1,k}^h,k+1,\ldots,\hat n\right),
\label{eq:BCF}
 \end{equation}
where a sum over the helicities $h$ of the intermediate gluon is implicit, and
\begin{eqnarray}
\label{eq:shift1} 
\hat{P}_{1,k}&=&P_{1,k}+\frac{P_{1,k}^2}{\langle n|P_{1,k}|1]}\lambda_n\tilde{\lambda}_1, \\ 
\label{eq:shift2}
\hat{p}_1&=&p_1+\frac{P_{1,k}^2}{\langle n|P_{1,k}|1]}\lambda_n\tilde{\lambda}_1,\\
\label{eq:shift3}
\hat{p}_n&=&p_n-\frac{P_{1,k}^2}{\langle n|P_{1,k}|1]}\lambda_n\tilde{\lambda}_1,
\end{eqnarray}
with $\lambda_i$ and $\tilde\lambda_i$ being the spinor components of $p_i=\lambda_i \, \tilde\lambda_i$.

As in Eq.~(\ref{eq:BCF}) we have to choose two reference gluons, $1$ and $n$, the color-flow decomposition and the color decomposition in the fundamental representation are not well suited to dress the BCF recursive relations with color, because they allow us to fix only one of the two reference gluons. 
The most natural color decomposition which fixes both reference gluons is the color decomposition in the adjoint representation, Eq.~(\ref{eq:adjoint_decomposition}). Inserting the color-ordered BCF relations, Eq.~(\ref{eq:BCF}), into Eq.~(\ref{eq:adjoint_decomposition}), one finds
\begin{eqnarray}
\begin{cal}A\end{cal}_n\left(1,\ldots,n\right) & = & \sum_{k=2}^{n-2}\sum_{\sigma \in  S_{n-2}}\left(F^{a_{\sigma_2}}\ldots F^{a_{\sigma_{n-1}}}\right)_{a_1a_n}A_{k+1}\left(\hat{1},\sigma_2,\ldots,\sigma_k,-\hat{P}_{1,\sigma_k}^{-h}\right)\nonumber \\
 &  & \qquad \frac{1}{P_{1,\sigma_k}^2}\ A_{n-k+1}\left(\hat{P}_{1,\sigma_k}^h, \sigma_{k+1},\ldots,\sigma_{n-1},\hat n\right), 
 \label{eq:AdjointBCF}
\end{eqnarray}
where 
\begin{equation}
P_{1,\sigma_k}=p_1+p_{\sigma_2}+\ldots+p_{\sigma_k}.
\end{equation}
For a given value of $k$, the sum over permutations appearing in Eq.~(\ref{eq:AdjointBCF}) can be decomposed in a similar way as for the three-gluon vertex part in the Berends-Giele recursive relations. The procedure is as follows
\begin{itemize}
\item[-] Choose an ordered partition $\pi=(\pi_1,\pi_2)$ of $\{2,3,\ldots,n-2,n-1\}$ such that $\#\pi_1=k-1$.
\item[-] Fix the first $k-1$ elements of the permutation to be in the subset $\pi_1$.
\item[-] Sum over all permutations of the first $k-1$ elements and over all permutations of the last $n-k-1$ elements.
\item[-] Sum over all possible choices for the ordered partition $\pi=(\pi_1,\pi_2)$.
\end{itemize}
This is equivalent to the replacement
\begin{equation}
\sum_{k=2}^{n-2}\ \sum_{\sigma \in S_{n-2}}\rightarrow \sum_{\pi \in OP(n-2,2)}\ \sum_{\sigma \in S_{k-1}} \ \sum_{\sigma ' \in S_{n-k-1}},
\label{eq:OPBCF}
\end{equation} 
where by $OP(n-2,2)$ we denote the set of all ordered partitions of $\{2,3,\ldots,n-1\}$.\\
Furthermore, for a fixed value of $k$, the color factor can be written
\begin{equation}
\left(F^{a_{\sigma_2}}\ldots F^{a_{\sigma_{n-1}}}\right)_{a_1a_n}=\left(F^{a_{\sigma_2}}\ldots F^{a_{\sigma_k}}\right)_{a_1x}\left(F^{a_{\sigma_{k+1}}}\ldots F^{a_{\sigma_{n-1}}}\right)_{xa_n},
\end{equation}
where a sum over $x=1,\ldots,8$ is understood.\\
Finally, the propagator clearly only depends on the choice of the ordered partition $\pi=(\pi_1,\pi_2)$ and not on the order of the elements in $\pi_1$ and $\pi_2$. If $\pi_1=\{\pi^2,\pi^3,\ldots,\pi^k\}$, we define
\begin{equation}
\begin{split}
P_{1,\pi_1} & = p_1+p_{\pi^2}+p_{\pi^3}+\ldots+p_{\pi^k},\\
P_{\pi_2,n} & = -P_{1,\pi_1}.
\end{split}
\end{equation}
At this point it is possible to identify subamplitudes in the expression for $\begin{cal}A\end{cal}_n$, namely
\begin{eqnarray}
\sum_{\sigma \in S_{k-1}}\left(F^{a_{\sigma_{\pi^2}}}\ldots F^{a_{\sigma_{\pi^k}}}\right)_{a_1x}A_{k+1}\left(\hat{1},\sigma_{\pi^2},\ldots,\sigma_{\pi^k},-\hat{P}_{1,\pi_1}^{-h}\right)\nonumber \\
 =\begin{cal}A\end{cal}_{k+1}\left(\hat{1},\pi_1,-\hat{P}_{1,\pi_1}^{-h,x}\right),
\end{eqnarray}
\begin{eqnarray}
\sum_{\sigma '\in S_{n-k-1}}\left(F^{a_{\sigma '_{\pi^{k+1}}}}\ldots F^{a_{\sigma '_{\pi^{n-1}}}}\right)_{xa_n}A_{n-k+1}\left(-\hat{P}_{\pi_2,n}^{-h},\sigma '_{\pi^{k+1}},\ldots,\sigma '_{\pi^{n-1}},\hat{n}\right)\nonumber\\
=\begin{cal}A\end{cal}_{n-k+1}\left(-\hat{P}_{{\pi_2},n}^{-h,x},\pi_2,\hat{n}\right),
\end{eqnarray}
where $x$ is the color of the intermediate gluon.\\
Collecting all the pieces, the color-dressed BCF recursive relations read
\begin{equation}
\begin{cal}A\end{cal}_n(1,2,\ldots,n)=\sum_{\pi \in  OP(n-2,2)}\begin{cal}A\end{cal}_{k+1}\left(\hat{1},\pi_1,-\hat{P}_{1,\pi_1}^{-h,x}\right)\frac{1}{P_{1,\pi_1}^2}\ \begin{cal}A\end{cal}_{n-k+1}\left(\hat{P}_{1,\pi_1}^{h,x},\pi_2,\hat{n}\right).
\label{eq:CDBCF}
\end{equation}
\myfigure{t!}{
\fcolorbox{white}{white}{
  \begin{picture}(363,162) (150,-61)
    \SetWidth{0.5}
    \SetColor{Black}
    \Text(303,19)[lc]{\Large{\Black{$=$}}}
    \SetWidth{0.5}
    \Line(210,19)(165,48)
    \Line(210,19)(165,-10)
    \Line(210,19)(257,48)
    \Line(210,19)(257,-10)
    \GOval(210,19)(30,30)(0){0.882}
    \Text(201,19)[lc]{\Large{\Black{$\begin{cal}A\end{cal}_n$}}}
    \Text(251,22)[lc]{\Large{\Black{$\vdots$}}}
    \Text(152,45)[lb]{\Large{\Black{$1$}}}
    \Text(150,-16)[lb]{\Large{\Black{$n$}}}
    \Text(332,19)[lc]{\Huge{\Black{$\Sigma_{\pi}$}}}
    \Text(368,-41)[lb]{\Large{\Black{$\hat n$}}}
    \Text(482,77)[lb]{\Large{\Black{$\pi$}}}
    \Text(483,-42)[lb]{\Large{\Black{$\bar \pi$}}}
    \Text(444,19)[lc]{\Large{\Black{$P_{1,\pi}$}}}
    \Text(372,75)[lb]{\Large{\Black{$\hat 1$}}}
    \Text(472,75)[lb]{\Large{\Black{$\vdots$}}}
    \Line(439,78)(439,-38)
    \Line(381,78)(440,78)
    \Line(444,94)(477,94)
    \Line(444,65)(477,65)
    \Line(444,-24)(477,-24)
    \Line(380,-38)(439,-38)
    \GOval(439,78)(23,26)(0){0.882}
    \Text(472,-44)[lb]{\Large{\Black{$\vdots$}}}
    \Line(444,-54)(477,-54)
    \GOval(439,-38)(23,26)(0){0.882}
    \Text(419,-44)[lb]{\large{\Black{$\begin{cal}A\end{cal}_{n-k+1}$}}}
    \Text(425,72)[lb]{\large{\Black{$\begin{cal}A\end{cal}_{k+1}$}}}
  \end{picture}
}}{
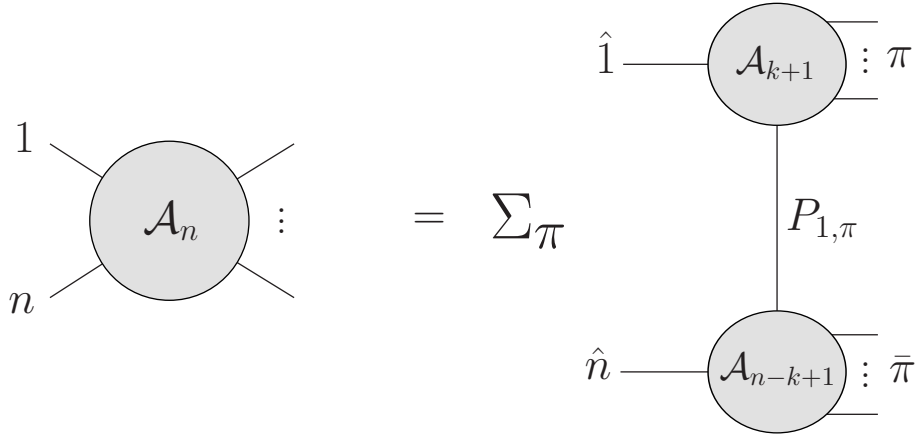
Diagrammatic representation of the color-dressed BCF recursive relations.
\label{fig:CDBCF}}

We emphasize that although the proof of these new recursive relations relies on the adjoint color basis, the final result, Eq.~(\ref{eq:CDBCF}), is independent of the choice of the basis. Furthermore, as in the case of the Berends-Giele recursive relations, we see that the form of the color-dressed BCF recursive relations stays the same as in the color-ordered case, with the only difference that in Eq.~(\ref{eq:CDBCF}) the sum goes over all partitions of $\{2,3,\ldots,n-1\}$, {\em i.e.}, over unordered objects. This implies that the new color-dressed BCF recursive relations have the same properties as in the color-ordered case, namely
\begin{enumerate}
\item The definition of the off-shell shifts, Eqs.~(\ref{eq:shift1}~-~\ref{eq:shift3}), is independent of the color.
\item As in the color-ordered BCF recursive relations, the pole structure of the scattering amplitude is manifest in Eq.~(\ref{eq:CDBCF}).
\item Similar to the color-ordered case, the subamplitudes in Eq.~(\ref{eq:CDBCF}) are not independent, but they are are linked via the off-shell shifts.
\end{enumerate}

The result~(\ref{eq:CDBCF}) obtained for amplitudes containing only gluons can be easily extended to include a single quark pair. For amplitudes containing a single $q\bar q$ pair, the color decomposition reads~\cite{Dixon:1996wi}
\begin{equation}
\begin{cal}A\end{cal}_n(1_q,2,\ldots,n-1,n_{\bar{q}})=\sum_{\sigma \in S_{n-2}}\left(T^{a_{\sigma_2}}\ldots T^{a_{\sigma_{n-1}}}\right)_i^{\phantom{i}\bar{\jmath}}\ A_n\left(1_q,\sigma_2,\ldots,\sigma_{n-1},n_{\bar{q}}\right).
\end{equation}
The BCF recursive relations for color-ordered amplitudes still hold when a quark pair is included, where either a quark or a gluon can be chosen as the intermediate particle~\cite{Luo:2005rx}. If a quark is chosen for the internal line, no sum over helicities has to be carried out, because helicity is conserved all along the fermion line. The BCF recursive relations then read 
\begin{align}\label{eq:quarkCDBCF}
A_n(1_q^h,2,\ldots,n_{\bar{q}}^{-h})=&\sum_{k=2}^{n-2}A_{k+1}\left(\hat{1}_q^h,2,\ldots,k,-\hat{P}_{\bar{q},1k}^{-h}\right)\\&\quad\quad\quad
\frac{1}{P_{q,1k}^2}A_{n-k+1}\left(\hat{P}_{q,1k}^h,k+1,\ldots,n-1,\hat n_{\bar{q}}^{-h}\right)\nonumber,
 \end{align}
where $h$ is the helicity of the quark.
Both the recursive relations and the color decomposition have the same form as in the case of a pure gluon amplitude, with the only difference that instead of working in the adjoint representation one now has to work in the fundamental representation of $SU(3)$. So the recursive relations derived in the case of pure gluon amplitudes can be easily extended to include a single $q\bar{q}$ pair
\begin{equation}
\begin{cal}A\end{cal}_n(1_q^h,2,\ldots,n_{\bar{q}}^{-h})=\sum_{\pi \in  OP(n-2,2)}\begin{cal}A\end{cal}_{k+1}\left(\hat{1}_q^{h},\pi_1,-\hat{P}_{\bar{q},1\pi_1}^{-h,x}\right)\frac{1}{P_{q,1\pi_1}^2}\ \begin{cal}A\end{cal}_{n-k+1}\left(\hat{P}_{q,1\pi_1}^{h,x},\pi_2,\hat{n}_{\bar{q}}^{-h}\right).
\end{equation}
The formula is exactly the same as in the pure gluon case, up to two small differences:
\begin{itemize}
\item[-] no helicity sum has to be carried out for the internal line
\item[-] $x$ is a color index in the fundamental representation.
\end{itemize}
It is possible to extend this relation to include photons, by simply performing the substitution
\begin{equation}
\left(T^a\right)_i^{\phantom{i}\bar{\jmath}}\;\to\;\delta_i^{\phantom{i}\bar{\jmath}}.
\end{equation}
From this it follows that a QED amplitude containing a single $q\bar q$ pair and $n-2$ photons can be written in terms of color-ordered amplitudes as
\begin{equation}
\begin{cal}A\end{cal}_n^{QED}(1_q,2,\ldots,n-1,n_{\bar q})=\sum_{\sigma \in S_{n-2}}A_n(1_q,\sigma_2,\ldots,\sigma_{n-1},n_{\bar q}).
\end{equation}
Thus the above recursive relations~(\ref{eq:quarkCDBCF}) still hold for QED amplitudes.
This particular result has already been pointed out by Stirling and Ozeren in Ref.~\cite{Ozeren:2005mp}. However, as shown there, for QED processes it is more efficient to take one of the fermions and one photon as reference particles. In fact, as there is no photon-photon vertex, all the terms in the recursive relations where both fermions are in the same subamplitude vanish, simplifying the calculation.

\section{Color-dressed CSW rules}
\label{sec:CDCSW}

\begin{fmffile}{cdrr_fg}
\newcommand{\dst}{\displaystyle}
\newcommand{\sst}{\scriptstyle}
\newcommand{\abr}[1]{\langle #1\rangle}
\newcommand{\spp}[1]{\left| #1\right.\rangle}
\newcommand{\mq}[1]{\text{``}#1\text{''}}
\newcommand{\ldu}{0.3}

\newcolumntype{x}[1]{>{\raggedleft}p{#1}} 
\newcolumntype{y}[1]{@{}p{#1}}

\fmfset{dot_len}{1mm}
\newcommand{\cswv}{\parbox{2cm}{\begin{center}
\begin{fmfgraph*}(25,20)
  \fmftop{i,j}
  \fmfbottom{ij}
  \fmf{plain}{i,v1,j}
  \fmf{plain,tension=1.5}{v1,ij}
  \fmffreeze
  \fmf{phantom}{v2,v1}
  \fmf{phantom,tension=5}{ij,v2}
  \fmfv{d.s=0,l.d=\ldu,l.a=90,l=$\sst i$}{i}
  \fmfv{d.s=0,l.d=\ldu,l.a=90,l=$\sst j$}{j}
  \fmfv{d.shape=circle,d.size=3,l.a=0,l=$\sst k$}{v2}
\end{fmfgraph*}\end{center}}}
\newcommand{\cswvp}{\parbox{2cm}{\begin{center}
\begin{fmfgraph*}(25,20)
  \fmftop{i,j}
  \fmfbottom{ij}
  \fmf{plain}{i,v1,j}
  \fmf{double,tension=1.5}{v1,v2}
  \fmf{phantom,tension=4.5}{v2,ij}
  \fmffreeze
  \fmfv{d.s=0,l.d=\ldu,l.a=90,l=$\sst i$}{i}
  \fmfv{d.s=0,l.d=\ldu,l.a=90,l=$\sst j$}{j}
  \fmfv{d.shape=circle,d.size=3,l.a=0,l=$\sst k$}{v2}
\end{fmfgraph*}\end{center}}}
\newcommand{\cswpvp}{\parbox{2cm}{\begin{center}
\begin{fmfgraph*}(25,20)
  \fmftop{i,j}
  \fmfbottom{ij}
  \fmf{plain}{v1,j}
  \fmf{double,tension=1.5}{v1,v2}
  \fmf{phantom,tension=4.5}{v2,ij}
  \fmf{double}{v1,i}
  \fmffreeze
  \fmfv{d.s=0,l.d=\ldu,l.a=90,l=$\sst i$}{i}
  \fmfv{d.s=0,l.d=\ldu,l.a=90,l=$\sst j$}{j}
  \fmfv{d.shape=circle,d.size=3,l.a=0,l=$\sst k$}{v2}
\end{fmfgraph*}\end{center}}}
\newcommand{\cswpv}{\parbox{2cm}{\begin{center}
\begin{fmfgraph*}(25,20)
  \fmftop{i,j}
  \fmfbottom{ij}
  \fmf{plain}{v1,j}
  \fmf{plain,tension=1.5}{v1,v2}
  \fmf{phantom,tension=4.5}{v2,ij}
  \fmf{double}{v1,i}
  \fmffreeze
  \fmfv{d.s=0,l.d=\ldu,l.a=90,l=$\sst i$}{i}
  \fmfv{d.s=0,l.d=\ldu,l.a=90,l=$\sst j$}{j}
  \fmfv{d.shape=circle,d.size=3,l.a=0,l=$\sst k$}{v2}
\end{fmfgraph*}\end{center}}}
\newcommand{\lcswpvp}{\parbox{2cm}{\begin{center}
\begin{fmfgraph*}(25,20)
  \fmftop{i,j}
  \fmfbottom{ij}
  \fmf{plain}{v1,i}
  \fmf{double,tension=1.5}{v1,v2}
  \fmf{phantom,tension=4.5}{v2,ij}
  \fmf{double}{j,v1}
  \fmffreeze
  \fmfv{d.s=0,l.d=\ldu,l.a=90,l=$\sst i$}{i}
  \fmfv{d.s=0,l.d=\ldu,l.a=90,l=$\sst j$}{j}
  \fmfv{d.shape=circle,d.size=3,l.a=0,l=$\sst k$}{v2}
\end{fmfgraph*}\end{center}}}
\newcommand{\lcswpv}{\parbox{2cm}{\begin{center}
\begin{fmfgraph*}(25,20)
  \fmftop{i,j}
  \fmfbottom{ij}
  \fmf{plain}{v1,i}
  \fmf{plain,tension=1.5}{v1,v2}
  \fmf{phantom,tension=4.5}{v2,ij}
  \fmf{double}{v1,j}
  \fmffreeze
  \fmfv{d.s=0,l.d=\ldu,l.a=90,l=$\sst i$}{i}
  \fmfv{d.s=0,l.d=\ldu,l.a=90,l=$\sst j$}{j}
  \fmfv{d.shape=circle,d.size=3,l.a=0,l=$\sst k$}{v2}
\end{fmfgraph*}\end{center}}}

In this section we present the color-dressing of the CSW vertex rules introduced in Refs.~\cite{Witten:2003nn,Cachazo:2005ga,Cachazo:2004kj}. 
These rules state that it is 
possible to build all color-ordered amplitudes from MHV vertices,
connecting them by scalar propagators. Each off-shell leg with momentum $P$ 
then corresponds to a spinor $P_{a\dot a}\eta^{\dot a}$, using some 
arbitrary antiholomorphic reference spinor $\eta^{\dot a}$. 
However, the CSW rules imply that an $n$-point color-ordered amplitude may have contributions from MHV vertices involving up to $n$ particles. The number of different vertices is thus growing steadily with the number of particles, which makes it impossible to put the Berends-Giele recursive relations and the CSW rules on the same footing. This problem has also been addressed in Ref.~\cite{Bena:2004ry}. 
Here we introduce a new method to decompose an MHV vertex into three-point vertices involving an auxiliary particle and derive recursive relations for both the auxiliary particle and the scalar propagators. Once these relations have been obtained, the corresponding color dressing is performed.

In analogy to the Berends-Giele recursive relations we define an $n$-point scalar 
off-shell current $J_h(1,\ldots,n)$ as the sum of all MHV diagrams with 
$n$ external on-shell legs, and one off-shell leg with helicity 
$h$.\footnote{In the context of the CSW rules, it makes sense to talk 
  about the helicity of an off-shell particle. As the off-shell continuation 
  of the spinors involves an arbitrary reference spinor $\eta^{\dot a}$, 
  the scalar off-shell currents are not gauge invariant objects. 
  However, the $\eta^{\dot a}$ dependence drops out in the end 
  \cite{Cachazo:2004kj}.} 
This scalar off-shell current can be easily constructed employing 
the CSW rules:
\begin{eqnarray}
J_h(1,\ldots,n)=\frac{1}{P_{1,n}^2} & \Bigg[ & \sum_{i=1}^{n-1}A_3\left(-P_{1,n}^{-h},\,P_{1,k}^{h_1},\,P_{k+1,n}^{h_2}\right)J_{h_1}(1,\ldots,i)\,J_{h_2}(i+1,\ldots,n)\nonumber\\  
& + & \sum_{i=1}^{n-2}\sum_{j=i+1}^{n-1}A_4\left(-P_{1,n}^{-h},\,P_{1,i}^{h_1},\,P_{i+1,j}^{h_2},\,P_{j+1,n}^{h_3}\right)J_{h_1}(1,\ldots,i)\nonumber\\ 
& & \qquad \qquad J_{h_2}(i+1,\ldots,j) J_{h_3}(j+1,\ldots,n)+\ldots\Bigg]\;,
\end{eqnarray}
where the dots indicate terms with higher order MHV vertices. A sum over helicities 
$(h,h_1,h_2,\ldots)$ with $-h+h_1+h_2+\ldots=n-4$ is implicitly
understood. According to the CSW rules, the vertices $A_n$ correspond to 
off-shell continued $n$-point MHV amplitudes.

However, as mentioned above in this form the CSW relations imply a factorial growth
in the color-dressed case, because of possibly large numbers of legs at 
single MHV vertices and the associated permutations of these legs. 
In order to tame this growth, we rewrite the CSW relations in a form 
similar to the Berends-Giele recursion with a tensor particle, 
where the terms in Table~\ref{tab:csw_buiding_blocks} serve as basic 
building blocks. We first introduce auxiliary double-lines, 
carrying threefold information:
\begin{itemize}
\item[-] The total momentum $P$ flowing in the double-line.
\item[-] A pair $(k_l,k_r)$, describing the momenta flowing in each 
  of the two lines separately. Notice that in general $k_l+k_r\neq P $.
\item[-] A pair $(a,b)$, describing the momenta of the negative helicity 
  legs in the corresponding MHV amplitude contained in the off-shell current. 
  If no negative helicity gluon is attached to the double-line, then $a=b=0$, 
  and if only one negative helicity gluon  is attached, then $b=0$. 
\end{itemize}
In order to build recursive relations where all $n$-point MHV vertices for 
$n\ge4$ are decomposed into three-point vertices, we define $n$-point 
(off-shell) double-line currents $J_{uv}^{ab}(1,\ldots,n)$, where $1\le u\le v<n$, as the sum of all diagrams 
with $n$ external on-shell legs and an (off-shell) auxiliary double-line.
This line carries the information $(k_l,k_r)=(P_{1,u},P_{v+1,n})$ and $(a,b)$, $a$ and $b$ being the momenta of the negative helicity gluons attached to it.\footnote{It is easy to see that all other assignments for $(k_l,k_r)$ do not contribute in the color-ordered case.} Furthermore, in the color-ordered case, $a$ must be of the form $a=P_{i,j}$ with $(i,j)$ constrained to one of the following possibilities
\begin{equation}
\begin{split}
1\le i & \le  j\le u,\\
u+1\le i & \le  j \le v,\\
v+1\le i & \le  j \le n
\end{split}
\end{equation}
and equivalently for $b$. Notice that according to our definition there are no one-point double-line currents. For later convenience we define all one-point double-line currents as zero.
All other double-line currents can be built recursively employing the vertices 
given in Table~\ref{tab:csw_buiding_blocks}, yielding
\begin{eqnarray}
 \label{eq:RecCOdouble}
J_{uv}^{ab}(1,\ldots,n) & = & \delta_{uv}\; V_{AG}^{h_1,h_2,ab}(P_{1,u},P_{u+1,n})\,J_{h_1}(1,\ldots,u)\,J_{h_2}(u+1,\ldots,n)\\
 & + &(1-\delta_{uv})\sum_{w=u}^{v-1} V_{AAR}^{a'b',h,ab}(P_{1,u},P_{w+1,v}P_{v+1,n})\,J_{uw}^{a'b'}(1,\ldots,v)\,J_h(v+1,\ldots,n),\nonumber
\end{eqnarray}
where sums over repeated indices are always understood.
\mytable{t!}{
\begin{tabular}{|c|r@{\hspace{1mm}}l|}
  \hline
  \multicolumn{3}{|c|}{\vphantom{\Large P}Right-handed Vertices}  \\ 
  \hline
  \cswv & $\dst V_{GG}^{h_i, h_j, h_k}(i,j,k)=$ & 
  $\dst\frac{\abr{\alpha\beta}^4}{\abr{ij}\abr{jk}\abr{ki}}$ \\
  \cswvp & $\dst V_{AG}^{h_i, h_j, a_kb_k}(i,j)=$ & 
  $\dst\frac{1}{\abr{ij}}\;\delta_{k_li}\delta_{k_rj}\;\epsilon(h_i,h_j,a_kb_k)$ \\
  \cswpvp & $\dst V_{AAR}^{a_ib_i, h_j, a_kb_k}(i_l,i_r,j)=$ & 
  $\dst\frac{1}{\abr{i_r j}}\;\delta_{k_li_l}\delta_{k_rj}\;\epsilon(a_ib_i,h_j,a_kb_k)$\\
  \cswpv & $\dst V_{GAR}^{a_ib_i, h_j, h_k}(i_l,i_r,j)=$ & 
  $\dst\frac{\abr{\alpha\beta}^4}{\abr{i_r j}\abr{jk}\abr{ki_l}}\;\epsilon(a_ib_i,h_j,h_k)$ \\
  \hline
\end{tabular}}{Right-handed basic building blocks in the MHV decomposition of color-ordered
  amplitudes in the CSW approach. 
  Details are given in the text.\label{tab:csw_buiding_blocks}}

The indices $\alpha$ and $\beta$ in Table~\ref{tab:csw_buiding_blocks} refer to
the two particles with negative helicity within one MHV amplitude. 
The $\epsilon$ functions appearing in the vertices involving an auxiliary double-line 
keep track of the negative helicity gluons attached to them,  
\begin{equation}
\epsilon \equiv\left\{\begin{array}{ll}
1, & \textrm{if }\# \textrm{negative helicity gluons } \le 2\\
0, & \textrm{if }\# \textrm{negative helicity gluons } > 2
\end{array}\right.\;.
\label{eq:eps}
\end{equation}
A vertex vanishes if the number of negative helicity gluons attached to a double-line is greater than two, since this situation corresponds 
to a non-MHV amplitude, that has to be decomposed further by means 
of the CSW relations.
No permutation of the incoming legs is allowed, since it would lead to 
double-counting. The respective rules in the $\overline{\text{MHV}}$ 
decomposition are obtained from the above by swaping helicities and 
replacing angular by square brackets.
In analogy to the double-line current, we can write the recursive relation for the scalar 
off-shell current in terms of double-line and scalar off-shell currents 
\begin{eqnarray}
 \label{eq:RecCOgluon}
&&\text{\hspace*{-10ex}}J_h(1,\ldots,n) = \frac{1}{P_{1,n}^2}  \sum_{k=1}^{n-1}\; \Bigg[ \;V_{GG}^{h_1,h_2,h}(P_{1,k},P_{k+1,n})\,J_{h_1}(1,\ldots,k)\,J_{h_2}(k+1,\ldots,n)\\
 && +\; \sum_{u=1}^{k-1} \sum_{v=u}^{k-1}
  V_{GAR}^{ab,h_{1},h}(P_{1,u},P_{v+1,k},P_{k+1,n})\,J_{uv}^{ab}(1,\ldots,k)\,J_{h_1}(k+1,\ldots,n)\Bigg]\nonumber.
 \end{eqnarray}
Notice that the second term vanishes for $k=1$ due to the vanishing of all one-point double-line currents.
 
The vertices given in Table~\ref{tab:csw_buiding_blocks} correspond to the situation where all gluons are attached on one side of the double-line. We will refer to these vertices as the \emph{right-handed vertices}. The right-handed vertices are sufficient to construct all MHV amplitudes. 
However, it is convenient for the subsequent color dressing of the CSW rules 
to recast Eq.~\eqref{eq:RecCOgluon} into a symmetric form. To do so, first we define left-handed vertices where all the gluons are attached to the opposite side of the double-line. These vertices are shown in Table~\ref{tab:csw_buiding_blocks2}.\footnote{As $V_{GG}$ and $V_{AG}$ are the same in both the right and left-handed decompositions, we do not list them again in Table~\ref{tab:csw_buiding_blocks2}.}
In the left-handed decomposition, the recursive relations Eq.~(\ref{eq:RecCOdouble}) and Eq.~(\ref{eq:RecCOgluon}) read
\begin{eqnarray}
&&\text{\hspace*{-5ex}}J_{uv}^{ab}(1,\ldots,n) = \delta_{uv}\; V_{AG}^{h_1,h_2,ab}(P_{1,u},P_{u+1,n})\,J_{h_1}(1,\ldots,u)\,J_{h_2}(u+1,\ldots,n)\\
 && + \;(1-\delta_{uv})\sum_{w=u+1}^{v} V_{AAL}^{h,a'b',ab}(P_{1,u},P_{u+1,w},P_{v+1,n})\,J_h(1,\ldots,u)\,J_{wv}^{a'b'}(u+1,\ldots,n),\nonumber
 \label{eq:LCOdouble}
\end{eqnarray}
\begin{eqnarray}
&&\text{\hspace*{-5ex}}J_h(1,\ldots,n) = \frac{1}{P_{1,n}^2}  \sum_{k=1}^{n-1}  \Bigg[ \; V_{GG}^{h_1,h_2,h}(P_{1,k},P_{k+1,n})\,J_{h_1}(1,\ldots,k)\,J_{h_2}(k+1,\ldots,n)\\
 && + \;  \sum_{u=1}^{n-k-1} \sum_{v=u}^{n-k-1}
  V_{GAL}^{h_{1},ab,h}(P_{1,k},P_{k+1,u},P_{v+1,n})\,J_{h_1}(1,\ldots,k)\,J_{uv}^{ab}(k+1,\ldots,n)\Bigg]\nonumber.
  \label{eq:LCOgluon}
 \end{eqnarray}
\mytable{t!}{
\begin{tabular}{|c|r@{\hspace{1mm}}l|}
  \hline
  \multicolumn{3}{|c|}{\vphantom{\Large P}Left-handed Vertices}  \\ 
  \hline
  \lcswpvp & $\dst V_{AAL}^{h_i,a_jb_j, a_kb_k}(i,j_l,j_r)=$ & 
  $\dst\frac{1}{\abr{ij_l}}\;\delta_{k_li}\delta_{k_rj_r}\;\epsilon(h_i,a_jb_j,a_kb_k)$\\
  \lcswpv & $\dst V_{GAL}^{h_i,a_jb_j, h_k}(i,j_l,j_r)=$ & 
  $\dst\frac{\abr{\alpha\beta}^4}{\abr{j_r k}\abr{ki}\abr{ij_l}}\;\epsilon(h_i,a_jb_j,h_k)$ \\
  \hline
\end{tabular}}{
  Left-handed basic building blocks in the MHV decomposition of color-ordered
  amplitudes in the CSW approach. 
  \label{tab:csw_buiding_blocks2}}
  
Combining the right- and left-handed decompositions, it is possible to write the recursive relations in a symmetric form involving both right-handed and left-handed vertices
\begin{eqnarray}
 \label{eq:SCOdouble}
&&\text{\hspace*{-5ex}}J_{uv}^{ab}(1,\ldots,n) = \delta_{uv}\; V_{AG}^{h_1,h_2,ab}(P_{1,u},P_{u+1,n})\,J_{h_1}(1,\ldots,u)\,J_{h_2}(u+1,\ldots,n)\\
&& + \;(1-\delta_{uv})\frac{1}{2}\sum_{w=u}^{v-1} V_{AAR}^{a'b',h,ab}(P_{1,u},P_{w+1,v},P_{v+1,n})\,J_{uw}^{a'b'}(1,\ldots,v)\,J_h(v+1,\ldots,n)\nonumber \\
 && + \;(1-\delta_{uv})\frac{1}{2}\sum_{w=u+1}^{v} V_{AAL}^{h,a'b',ab}(P_{1,u},P_{u+1,w},P_{v+1,n})\,J_h(1,\ldots,u)\,J_{wv}^{a'b'}(u+1,\ldots,n)\nonumber, 
\end{eqnarray}
\begin{eqnarray} \label{eq:SCOgluon}
&&\text{\hspace*{-5ex}}J_h(1,\ldots,n) = \frac{1}{P_{1,n}^2}  \sum_{k=1}^{n-1}  \Bigg[\; V_{GG}^{h_1,h_2,h}(P_{1,k},P_{k+1,n})\,J_{h_1}(1,\ldots,k)\,J_{h_2}(k+1,\ldots,n)\\
&& + \; \frac{1}{2} \sum_{u=1}^{k-1} \sum_{v=u}^{k-1}
  V_{GAR}^{ab,h_{1},h}(P_{1,u},P_{v+1,k},P_{k+1,n})\,J_{uv}^{ab}(1,\ldots,k)\,J_{h_1}(k+1,\ldots,n)\nonumber \\
 && + \; \frac{1}{2} \sum_{u=1}^{n-k-1} \sum_{v=u}^{n-k-1}
  V_{GAL}^{h_{1},ab,h}(P_{1,k},P_{k+1,u},P_{v+1,n})\,J_{h_1}(1,\ldots,k)\,J_{uv}^{ab}(k+1,\ldots,n)\Bigg]\nonumber.
 \end{eqnarray}
These recursive relations, equivalent to the CSW vertex rules, can be solved simultaneously to construct the scalar off-shell current. The difference to the pure CSW approach without decomposition of the MHV vertices lies in the fact that the number of different vertices in Eq.~(\ref{eq:SCOdouble}) and Eq.~(\ref{eq:SCOgluon}) is fixed and does not grow with the number of particles. 
The approach thereby differs from the one presented in Ref.~\cite{Bena:2004ry}. Furthermore, as the number of different vertices is fixed and as only three-point vertices are present, these new recursive relations are well suited to be compared to the Berends-Giele recursive relations. 

We now turn to the color dressing of the new CSW-like recursive relations. 
The procedure is very similar to the Berends-Giele case, but it 
contains some technical subtleties. 
The interested reader may refer to  appendix A for a detailed discussion.
As the new recursive relations only contain three-point vertices, we expect the color-dressed vertices to be of the same form as in Eq.~(\ref{eq:CD3vertex}), 
\begin{equation}
\begin{cal}V\end{cal}(P,Q)=\delta_{L}^{\bar{G}}\delta_{N}^{\bar{K}}\delta_{H}^{\bar{M}}\ V_R(P,Q)+\delta_{N}^{\bar{G}}\delta_{L}^{\bar{M}}\delta_{H}^{\bar{K}}\ V_L(Q,P).
\label{eq:CD3vertexCSW}
\end{equation} 
In the color-flow basis, the scalar off-shell currents are defined in the usual way
\begin{equation}
\begin{cal}J\end{cal}_h^{I\bar J}(1,\ldots,n)  =  \sum_{\sigma \in S_n}\delta_{i_{\sigma_1}}^{\bar{J}}\delta_{i_{\sigma_2}}^{\bar{\jmath}_{\sigma_1}}\ldots\delta_{I}^{\bar{\jmath}_{\sigma_n}}J_h(\sigma_1,\sigma_2\ldots,\sigma_n).
\end{equation}
The color-dressed $n$-point double-line currents are defined by
\begin{eqnarray}
\label{eq:CDdoubleline}
\begin{cal}J\end{cal}_{\pi_l\pi_r}^{ab,I\bar J}(\pi_l,\pi_m,\pi_r) & = & \sum_{\sigma \in S_u}\sum_{\sigma' \in S_{v-u}}\sum_{\sigma'' \in S_{n-v}} \Bigg[  \delta_{i_{\sigma_{l,1}}}^{\bar{J}}\ldots\delta_{i_{\sigma'_{m,u+1}}}^{\bar{\jmath}_{\sigma_{l,u}}}\ldots 
\delta_{i_{\sigma''_{r,{v+1}}}}^{\bar{\jmath}_{\sigma'_{m,{v}}}}\ldots\delta_{I}^{\bar{\jmath}_{\sigma''_{r,n}}}\\
 & & \qquad J_{uv}^{ab}(\sigma_{\pi_l},\sigma'_{\pi_m},\sigma''_{\pi_r})
  + 
\delta_{i_{\sigma''_{r,{v+1}}}}^{\bar{J}}\ldots\delta_{i_{\sigma'_{m,{u+1}}}}^{\bar{\jmath}_{\sigma''_{r,n}}}\ldots\delta_{i_{\sigma_{l,1}}}^{\bar{\jmath}_{\sigma'_{m,{v}}}}\ldots\delta_{I}^{\bar{\jmath}_{\sigma_{l,u}}}\nonumber \\
 & & \qquad \qquad J_{(n-v)(n-u)}^{ab}(\sigma''_{\pi_r},\sigma'_{\pi_m},\sigma_{\pi_l})
\Bigg]\nonumber,
\end{eqnarray}
where $\pi_l$ and $\pi_r$ are two proper subsets of $\{1,2,\ldots,n\}$ referring to the momenta flowing in each  line separately, $(k_l,k_r)=(P_{\pi_l},P_{\pi_r})$, and $\pi_m$ is defined by $\pi_l\oplus\pi_m\oplus\pi_r=\{1,2,\ldots,n\}$ (Notice that $\pi_m$ may be empty). On the right-hand side of Eq.~(\ref{eq:CDdoubleline}), the indices $(u,v)$ of the color-ordered currents are defined by $(u,v)=(\#\pi_l,n-\#\pi_r)$ and
\begin{equation}
\begin{split}
\pi_l & =  \{\pi_l^1,\pi_l^2,\ldots,\pi_l^u\}, \\
\pi_m & =  \{\pi_m^{u+1},\pi_m^{u+2},\ldots,\pi_m^v\}, \\
\pi_r & =  \{\pi_r^{v+1},\pi_r^{v+2},\ldots,\pi_r^n\}. 
\end{split}
\end{equation}
Finally, the symbols $\sigma_{i,j}$ are defined by $\sigma_{i,j}=\sigma(\pi_i^j)$. Notice that due to the second term appearing in Eq.~(\ref{eq:CDdoubleline}), a color-dressed $n$-point double-line current is symmetric in $(\pi_l,\pi_r)$.\\
The color dressing is similar to the Berends-Giele case and the result is
\begin{eqnarray}
\label{eq:CDCSW1}
\begin{cal}J\end{cal}_h^{I\bar{J}}(1,\ldots,n)=\frac{1}{P_{1,n}^2} & \Bigg[ & \sum_{\pi\in P(n,2)}\begin{cal}V\end{cal}_{GG}^{h_1,h_2,h}(P_{\pi_1},P_{\pi_2})\begin{cal}J\end{cal}_{h_1}^{K\bar{L}}(\pi_1)\begin{cal}J\end{cal}_{h_2}^{M\bar{N}}(\pi_2)\\
 & + & \frac{1}{2}\sum_{\pi\in OP(n,3)} \begin{cal}V\end{cal}_{GA}^{ab,h_1,h}(P_{\pi_1},P_{\pi_2},P_{\pi_3}) 
\begin{cal}J\end{cal}_{\pi_1\pi_2}^{ab,K\bar{L}}(\pi_1,\pi_2)\begin{cal}J\end{cal}_{h_1}^{M\bar{N}}(\pi_3)\nonumber \\
 & + & \frac{1}{2}\sum_{\pi\in OP(n,4)} \begin{cal}V\end{cal}_{GA}^{ab,h_1,h}(P_{\pi_1},P_{\pi_3},P_{\pi_4}) 
\begin{cal}J\end{cal}_{\pi_1\pi_3}^{ab,K\bar{L}}(\pi_1,\pi_2,\pi_3)\begin{cal}J\end{cal}_{h_1}^{M\bar{N}}(\pi_4)\Bigg ], \nonumber
\end{eqnarray}
\begin{eqnarray}
\label{eq:CDCSW2}
\begin{cal}J\end{cal}_{\pi_l,\pi_r}^{ab,I\bar{J}}(\pi_l,\pi_m,\pi_r)&=& \delta_{uv}\;\begin{cal}V\end{cal}_{AG}^{h_1,h_2,ab}(P_{\pi_l},P_{\pi_r})\begin{cal}J\end{cal}_{h_1}^{K\bar{L}}(\pi_l)\;\begin{cal}J\end{cal}_{h_2}^{M\bar{N}}(\pi_r)\\
 & + & (1-\delta_{uv})\frac{1}{2}\Bigg[ \begin{cal}V\end{cal}_{AA}^{a'b',h,ab}(P_{\pi_l},P_{\pi_m},P_{\pi_r})\;\begin{cal}J\end{cal}_{\pi_l\pi_2}^{a'b',K\bar{L}}(\pi_l,\pi_m)\;\begin{cal}J\end{cal}_h^{M\bar{N}}(\pi_r)\nonumber \\
  & + & \sum_{\pi\in OP(\pi_m,2)}
 \begin{cal}V\end{cal}_{AA}^{a'b',h,ab}(P_{\pi_l},P_{\pi_2},P_{\pi_r}) \;\begin{cal}J\end{cal}_{\pi_l\pi_2}^{a'b',K\bar{L}}(\pi_l,\pi_1,\pi_2)\;\begin{cal}J\end{cal}_h^{M\bar{N}}(\pi_r) \nonumber \\
  & + & (\pi_l \leftrightarrow \pi_r) \Bigg], \nonumber
\end{eqnarray}
where in Eq.~(\ref{eq:CDCSW2}) $OP(\pi_m,2)$ is the set of all ordered partitions of $\pi_m$ into two independent parts.\\
Apart from a few subtleties, the procedure of the color dressing is 
now similar to the Berends-Giele case. A detailed discussion is given in  
appendix A. The main differences are the following:
\begin{itemize}
\item[-] As the color-dressed vertices, Eq.~(\ref{eq:CD3vertexCSW}), have both right and left-handed contributions, the symmetric form for the color-ordered recursive relations, Eq.~(\ref{eq:SCOdouble}) and Eq.~(\ref{eq:SCOgluon}) is employed.
\item[-] The contributions coming from the $u=v$ and $u\neq v$ terms in Eq.~(\ref{eq:SCOdouble}) and Eq.~(\ref{eq:SCOgluon}) are treated separately. For example, the $u=v$ and $u\neq v$ contributions in Eq.~(\ref{eq:SCOgluon}) give rise to the $OP(n,3)$ and $OP(n,4)$ terms in Eq.~(\ref{eq:CDCSW1}), respectively.
\end{itemize}
Apart from these, the new CSW-like recursive relations retain the same form as the corresponding color-ordered relations with the difference that in the color-ordered case the sum goes over unordered objects. Furthermore, as in the color-ordered case, the number of different vertices is fixed and only three-point vertices appear in the recursive relations. Therefore we may compare them to the color-dressed Berends-Giele recursive relation presented in Section~\ref{sec:BG}.

\end{fmffile}

\section{Numerical results}
\label{sec:results}

\newcolumntype{x}[1]{@{}>{\raggedleft}p{#1}}
\newcolumntype{y}[1]{@{}p{#1}}

All relations for calculating multi-gluon amplitudes presented 
in the previous sections have been implemented into C++ Monte Carlo 
programs using the tools set {\tt ATOOLS-2.0} and the integration package
{\tt PHASIC++-1.0} \cite{Gleisberg:2003xi}. 
A comparison of calculation times for helicity summed color-ordered amplitudes 
versus the results obtained in Ref.~\cite{Dinsdale:2006sq} has been performed.
Our implementations yield exactly the same growth 
in computation time, except for the CSW rules, where
we gain considerably due to rewriting the CSW vertex rules in terms of 
recursive relations for internal lines.
Furthermore we have checked, employing the color-flow basis, that the 
color-dressed relations yield the same results as the calculations employing 
color-ordered amplitudes along with the color-flow decomposition
presented in Ref.~\cite{Maltoni:2002mq}.
Using the adjoint representation, we have checked that the 
color-dressed BCF relations yield the same result as the color-ordered
ones along with a decomposition of the total amplitude in the adjoint basis. 

\mytable{t!}{
\begin{tabular}{|l|
                 x{1.45cm}y{0.5cm}|x{1.45cm}y{0.5cm}|
                 x{1.45cm}y{0.5cm}|x{1.45cm}y{0.5cm}|
                 x{1.45cm}y{0.5cm}|x{1.45cm}y{0.5cm}|}
  \hline\vphantom{\Large P}
  \centering Final & \multicolumn{4}{c|}{BG} & \multicolumn{4}{c|}{BCF} &
  \multicolumn{4}{c|}{CSW} \\\cline{2-13}\vphantom{\Large P}
  \centering State & \multicolumn{2}{c|}{CO} & \multicolumn{2}{c|}{CD} & 
  \multicolumn{2}{c|}{CO} & \multicolumn{2}{c|}{CD} & 
  \multicolumn{2}{c|}{CO} & \multicolumn{2}{c|}{CD} \\
  \hline
  $2g$  & 0&.24    & 0&.28   & 0&.28   & 0&.33   & 0&.31    & 0&.26   \\
  $3g$  & 0&.45    & 0&.48   & 0&.42   & 0&.51   & 0&.57    & 0&.55   \\
  $4g$  & 1&.20    & 1&.04   & 0&.84   & 1&.32   & 1&.63    & 1&.75   \\
  $5g$  & 3&.78    & 2&.69   & 2&.59   & 7&.26   & 5&.95    & 5&.96   \\
  $6g$  & 14&.2    & 7&.19   & 11&.9   & 59&.1   & 27&.8    & 30&.6   \\
  $7g$  & 58&.5    & 23&.7   & 73&.6   & 646&    & 146&     & 195&    \\
  $8g$  & 276&     & 82&.1   & 597&    & 8690&   & 919&     & 1890&   \\
  $9g$  & 1450&    & 270&    & 5900&   & 127000& & 6310&    & 29700&  \\
  $10g$ & 7960&    & 864&    & 64000&  & -&      & 48900&   & -& \\
  \hline
\end{tabular}}{
  Computation time ($s$) of the $2\to n$ gluon amplitudes for $10^4$ 
  phase space points, sampled over helicity and color. Results are given for 
  the color-ordered (CO) and the color-dressed (CD) Berends-Giele (BG),
  Britto-Cachazo-Feng (BCF) and Cachazo-Svr\v{c}ek-Witten (CSW) relations.
  Numbers were generated on a 2.66~GHz Xeon{\texttrademark} CPU. 
  \label{tab:comparison_co_cd}}
A comparison of the computation times for the various approaches 
using the color-flow basis can be found in Table~\ref{tab:comparison_co_cd}.
The color-dressed Berends-Giele relations are the fastest 
method for more than five final state gluons. For less than six outgoing 
gluons the color-flow decomposition using color-ordered amplitudes calculated 
according to the BCF recursion performs better. In this case only few 
valid color flows exist \cite{Maltoni:2002mq} and primarily (or only) 
MHV vertices contribute. For those the computation time increases only 
linearly with the number of outgoing particles in the color-ordered BCF 
relations.

It is apparent that the computation times in the color-dressed BCF and 
in the color-dressed CSW case grow very fast. 
In the case of the CSW relations the reason is the number of types of 
internal lines, which is larger than in the Berends-Giele and in the 
BCF approach. 
In this respect it is important to note that each double line may eventually 
carry zero, one or two indices of attached negative helicity gluons.
Additionally, in most cases two vertices exist for either of these lines
(cf.\ Table~\ref{tab:csw_buiding_blocks}), 
yielding a large amount of lines that finally have to be computed. 
However, the growth we encounter by employing this method is still 
not factorial but exponential. Nevertheless the factor in the 
exponent is still too large for the method to be competitive with the 
Berends-Giele approach.
This fact is illustrated in Table~\ref{tab:comparison_nzt}, where we list
the average number of nonzero internal lines counted either by value or by 
origination vertex. The former corresponds to the average number of nonzero 
currents in the Berends-Giele approach. 
\mytable{h}{
\begin{tabular}{|c|x{1.25cm}y{0.75cm}|x{1.75cm}y{0.75cm}|x{1.75cm}y{0.75cm}|
                 x{1.75cm}y{0.75cm}|x{1.75cm}y{0.75cm}|}
  \hline\vphantom{\Large P}Final & \multicolumn{2}{c|}{Currents} & 
  \multicolumn{4}{c|}{Internal lines (CSW) by} &
  \multicolumn{4}{c|}{MHV vertices (BCF) by} \\\cline{4-11}
  \vphantom{\Large P}State & \multicolumn{2}{c|}{(BG)} &
  \multicolumn{2}{c|}{vertex} & \multicolumn{2}{c|}{value} & 
  \multicolumn{2}{c|}{vertex} & \multicolumn{2}{c|}{value} \\
  \hline
  $2g$ & 7&.04   & 3&.48   & 7&.56   & 1&.98   & 1&.98    \\
  $3g$ & 19&.50  & 8&.56   & 27&.45  & 4&.43   & 4&.57    \\
  $4g$ & 44&.67  & 18&.58  & 109&.0  & 14&.13  & 18&.17   \\
  $5g$ & 95&.74  & 38&.63  & 407&.4  & 63&.88  & 126&.3   \\
  $6g$ & 198&.8  & 78&.25  & 1648&   & 297&.2  & 1026&    \\
  $7g$ & 405&.8  & 157&.8  & 6773&   & 1395&   & 10330&   \\
  $8g$ & 850&.3  & 325&.8  & 31340&  & 6073&   & 124600&  \\
  \hline
\end{tabular}}{
   Average number of nonzero currents in the color-dressed
   Berends-Giele relations, average number of internal lines in the
   CSW approach and average number of nonzero MHV vertices 
   in the color-dressed BCF relations using the color-flow decomposition. 
   MHV vertices in BCF are counted either by distinct value 
   or by distinct assignment of unshifted external momenta. 
   Internal lines in CSW are counted either by vertex or by distinct value.
   \label{tab:comparison_nzt}}

Employing the color-dressed BCF relations, we encounter a factorial growth 
of the computation time. We have identified three main reasons:
\begin{itemize}
\item[-] The subamplitudes are linked by the spinor shifts.
\item[-] The natural color basis is the adjoint basis.
\item[-] The amplitudes are decomposed down to three-point vertices.
\end{itemize}
We address these points in order.
 
In the color-dressed as well as in the color-ordered BCF relations, 
Eqs.~\eqref{eq:BCF} and~\eqref{eq:CDBCF}, the subamplitudes of a given 
decomposition are linked via the shifts 
Eqs.~(\ref{eq:shift1}-\ref{eq:shift3}). Thus the BCF relations 
need a recursive calculation of subamplitudes in the sense that the 
total amplitude is to be decomposed successively into smaller building blocks, 
finally yielding only three-point MHV vertices. 
In other words, we have to take Eq.~\eqref{eq:CDBCF} literally
and apply a top-down approach of the computation, since for the evaluation
of each subamplitude all previous spinor shifts have to be computed.
Figuratively speaking, this is due to the fact that in the BCF recursion 
all subamplitudes ``remember'' which decomposition they originated from, 
thus inhibiting the calculation of general color-dressed subamplitudes.
This fact is also illustrated in Table~\ref{tab:comparison_nzt}, 
where we list the average number of distinct nonzero MHV vertices 
and the average number of distinct assignments of unshifted momenta 
at these vertices. 
The latter corresponds to the average number of internal lines in the 
CSW approach, counted by origination vertex. It grows much slower than 
the former, although faster than for example the average number of nonzero 
currents in the Berends-Giele relations. 

When applying the top-down procedure of the computation described above,
it is necessary to avoid the calculation of terms yielding zero due to 
the color assignment of external and internal lines.
This can be done in two steps.
First, all valid color flows are identified employing an algorithm 
similar to the one used for the Berends-Giele recursion. 
Second, the subamplitudes are calculated only for the valid color structures.
The calculation can be alleviated if the reference 
particles in the recursion are chosen such that together they form a color 
current having a nonvanishing contribution to the respective amplitude. 
Since there exists no decomposition assigning both particles to a common 
subamplitude, the corresponding color current does not contribute 
anymore. This procedure eliminates many terms in the recursion, but it is still
insufficient in the case of the color-flow basis. 
In fact we expect some redundancy in the calculation of color-ordered 
subamplitudes due to the dual Ward identities, which is introduced by 
fixing the reference particles for all possible color flows of an 
amplitude simultaneously, cf.\ Eq.~\eqref{eq:CDBCF}.
To see this, consider a dual Ward identity of the form
\begin{equation}\label{dwi_cdbcf}
  A(2,1,3,\ldots,n)=-\sum_{l\ne2;\;1\le l<n}A(1,\ldots,l,2,l+1,\ldots,n)\;.
\end{equation}
Assume that particles $1$ and $n$ have been fixed to be the reference particles
in the recursion and the ordering $\{2,1,3,\ldots,n\}$ yields a valid 
color flow. In this case the above choice of reference particles
is actually inconvenient to calculate the respective contribution to the 
total amplitude, since the sum on the right hand side of Eq.~\eqref{dwi_cdbcf}
could be replaced by the one term on the left hand side.
This problem does not occur in the color-ordered case, since the reference
particles are chosen separately for each color flow. To illustrate this, in 
Table~\ref{tab:comparison_nz_bcf_ard} we compare the ratio of the average 
number of distinct nonzero MHV vertices in the color-dressed and the 
color-ordered BCF relations for the color-flow basis and the 
adjoint representation incorporating all simplifications described above. 
In the adjoint representation the color-dressed relations 
yield less terms than the color-ordered ones, since the adjoint 
representation naturally avoids the problem of encountering singlet gluons, 
that decouple. However, much more effort is spent on the computation of 
color factors in the adjoint representation \cite{Maltoni:2002mq},
such that it is not the method of choice. 
\mytable{t}{
\begin{tabular}{|x{0.75cm}y{1.25cm}|x{1.25cm}y{1.25cm}|x{1.25cm}y{1.25cm}|}
  \hline\multicolumn{2}{|c|}{\vphantom{\Large P}Process} & 
  \multicolumn{2}{c|}{Color flow} & \multicolumn{2}{c|}{Adjoint} \\
  \hline
  $gg$&$\,\to 2g$ & 0&.78  & 0&.86  \\
  $gg$&$\,\to 3g$ & 0&.83  & 0&.74  \\
  $gg$&$\,\to 4g$ & 0&.94  & 0&.60  \\
  $gg$&$\,\to 5g$ & 1&.14  & 0&.51  \\
  $gg$&$\,\to 6g$ & 1&.44  & 0&.44  \\
  \hline
\end{tabular}}{
   Ratio of the average number of nonzero MHV vertices in the color
   dressed and the color-ordered case in the color-flow and the 
   adjoint representation decomposition.
   \label{tab:comparison_nz_bcf_ard}}

In the color-dressed BCF relations each amplitude is decomposed completely 
into three-point vertices. In contrast, in the color-ordered case, 
any MHV amplitude occuring in any step of the recursion can be evaluated 
immediately. To highlight the differences due to this treatment, 
Table~\ref{tab:comparison_tv_bcf} shows a comparison of 
the average number of distinct nonzero MHV vertices that have to be 
evaluated in the color-dressed and in the color-ordered case. 
We also give the same number for the color-ordered case, 
when each amplitude is decomposed into three-point vertices as well.
\mytable{h}{
\begin{tabular}{|x{0.75cm}y{1.25cm}|
                 x{1.75cm}y{0.5cm}|x{1.75cm}y{0.5cm}|
                 x{1.75cm}y{0.75cm}|}
  \hline
  \multicolumn{2}{|c|}{\vphantom{\Large P}Process} & 
  \multicolumn{4}{c|}{CO} & \multicolumn{2}{c|}{CD} \\
  \cline{3-6}\multicolumn{2}{|c|}{\vphantom{\Large P}} & 
  \multicolumn{2}{c|}{general MHV} & \multicolumn{2}{c|}{3-point MHV} &&\\
  \hline
  $gg$&$\,\to 2g$ & 1&.28    & 2&.55   & 1&.98    \\
  $gg$&$\,\to 3g$ & 1&.84    & 5&.51   & 4&.57    \\
  $gg$&$\,\to 4g$ & 7&.41    & 19&.33  & 18&.17   \\
  $gg$&$\,\to 5g$ & 48&.78   & 110&.7  & 126&.3   \\
  $gg$&$\,\to 6g$ & 318&.3   & 714&.7  & 1026&    \\
  $gg$&$\,\to 7g$ & 2329&    & 5269&   & 10330&   \\
  $gg$&$\,\to 8g$ & 20650&   & 46890&  & 124600&  \\
  \hline
\end{tabular}}{
   Average number of nonzero MHV vertices in the color-flow decomposition
   for the color-ordered (CO) and the color-dressed (CD) BCF relations.
   \label{tab:comparison_tv_bcf}}\\

\section{Conclusions}
\label{sec:conclusions}

We have presented a new approach to the calculation of multi-parton
amplitudes which extends the recursive relations for the color-ordered
amplitudes to relations for the full colored amplitudes.  We have
argued that in general these new color-dressed relations should be
more suitable for a numerical implementation since they naturally
avoid the factorial growth implicit in taking the sum over the
permutations of the possible color flows in an amplitude.  The taming
of the factorial growth to an exponential one is easily proved in the
color-dressed formulation of the Berends-Giele recursive relations
which we find to be the same as the Schwinger-Dyson approach introduced
in Ref.~\cite{Draggiotis:2002hm} and equivalent to the ALPHA algorithm
of Ref.~\cite{Caravaglios:1998yr}.  Using a similar approach but
exploiting the adjoint color basis decomposition, Eq.~(\ref{eq:adjoint_decomposition}),
we have proved a new formulation of the BCF relations,
Eq.~(\ref{eq:CDBCF}), which involves the full amplitudes,
including color, and retains the same formal simplicity of the
original formulation. Finally, we have considered the CSW relations.
In this case we had first to recast them in a form similar to the
Berends-Giele relation through the introduction of a new type of
three-point vertices and effective particles, Eqs.~(\ref{eq:SCOdouble},\ref{eq:SCOgluon}).  
It is interesting to note that while for the Berends-Giele
relations the color dressing is straightforward due to the close
correspondence to the Feynman diagram approach, this is far less
trivial for the BCF and CSW relations, for which there is no direct
relation to the standard quantum field theory perturbative approach.

To test the numerical efficiency of the different formulations we have
also implemented the corresponding algorithms and computed squared
amplitudes for $2\to n$ gluon scattering, by performing the sum over
helicities and color with a Monte Carlo method. Our results clearly
show the numerical superiority of the recursive formulation by 
Berends and Giele over all twistor-inspired methods, both from the 
point of view of the growth of complexity with $n$ and the simplicity of the
implementation. For the color-ordered amplitude formulation we confirm
the results of Ref.~\cite{Dinsdale:2006sq} except for the CSW
relations, on which we improve considerably by bringing them to the
same level of complexity as the BCF relations.  The color-dressed
formulations of the BCF and CSW relations perform worse than the
corresponding color-dressed Berends-Giele relations for different
reasons. The BCF relations are penalized by their top-down structure,
{\it i.e.}, the fact that for each helicity and color configuration
the decomposition in terms of amplitudes with smaller multiplicity has
to be found, and from the fact that their natural (and minimal)
color basis is the adjoint basis which is computationally quite heavy. The
``improved'' color-dressed CSW relations instead suffer from the
presence of a large number of elementary line types and effective three-point
vertices which eventually affect the overall growth of the algorithm.

In conclusion, we have shown how color can be included in the
color-stripped recursive relations coming from twistor-inspired
methods that do not have a straightforward relation with a standard
perturbative Lagrangian approach. The resulting color-dressed BCF
relations can be easily derived by employing the adjoint color decomposition, 
which exactly matches onto the BCF structure, and retain the same very
elegant form of their color-ordered counterpart.  In this respect, it
is suggestive to speculate that similar colored relations might be derived
also for one-loop amplitudes, for which an analogous color decomposition
holds and may be deduced from an effective QCD Lagrangian, still unknown.

\vskip 2 ex 
\noindent
{\large\bf Acknowledgments}

\noindent
We like to thank Johan Alwall, Vittorio del Duca and Rikkert Frederix for
their comments on the manuscript.\\[0.5ex]
This work was supported by the Belgian Federal Office for Scientific, 
Technical and Cultural Affairs through the 
Interuniversity Attraction Pole P5/27. 

\appendix
\section{Appendix}
In this appendix we give a detailed derivation for the color dressing of the CSW-like recursive relations presented in Section~\ref{sec:CDCSW}.
We start with the color dressing of the recursive relations for the double-line current. As we want all color-dressed vertices to be of the form given in Eq.~(\ref{eq:CD3vertexCSW}), involving both right-handed and left-handed contributions, we use the symmetric form of the recursive relation, Eq.~(\ref{eq:SCOdouble}). We will consider the terms for $u=v$ and $u\neq v$ appearing in Eq.~(\ref{eq:SCOdouble}) separately and we will start with the $u=v$ term. This term corresponds to $\pi_m=\emptyset$. Inserting it into the definition~(\ref{eq:CDdoubleline}), we obtain
\begin{eqnarray}\label{eq:CDUeqV}
\sum_{\sigma \in S_u}\sum_{\sigma'' \in S_{n-u}} & \Bigg[ &  \delta_{i_{\sigma_{l,1}}}^{\bar{J}}\ldots\delta^{\bar{\jmath}_{\sigma_{l,u}}}_{i_{\sigma''_{r,{u+1}}}}\ldots\delta_{I}^{\bar{\jmath}_{\sigma''_{r,n}}}V_{AG}^{h_1,h_2,ab}(P_{\sigma_{\pi_l}},P_{\sigma''_{\pi_r}})\,J_{h_1}(\sigma_{\pi_l})\,J_{h_2}(\sigma''_{\pi_r})\\
 & + &
\delta_{i_{\sigma''_{r,{u+1}}}}^{\bar{J}}\ldots\delta^{\bar{\jmath}_{\sigma''_{r,n}}}_{i_{\sigma_{l,1}}}\ldots\delta_{I}^{\bar{\jmath}_{\sigma_{l,u}}}
 V_{AG}^{h_2,h_1,ab}(P_{\sigma''_{\pi_r}},P_{\sigma_{\pi_l}})\,J_{h_2}(\sigma''_{\pi_r})\,J_{h_1}(\sigma_{\pi_l})
\Bigg]\nonumber.
\end{eqnarray}
The decomposition of the two color factors is similar to the Berends-Giele case, Eq.~(\ref{eq:ColorFactDecomp3V}),
\begin{eqnarray}
\delta_{i_{\sigma_{l,1}}}^{\bar{J}}\ldots\delta^{\bar{\jmath}_{\sigma_{l,u}}}_{i_{\sigma''_{r,{u+1}}}}\ldots\delta_{I}^{\bar{\jmath}_{\sigma''_{r,n}}} & = & \delta_L^{\bar{J}}\delta_N^{\bar{K}}\delta_I^{\bar{M}}\quad \delta_{i_{\sigma_{l,1}}}^{\bar{L}}\ldots\delta^{\bar{\jmath}_{\sigma_{l,u}}}_{K}\quad \delta^{\bar{N}}_{i_{\sigma''_{r,{u+1}}}}\ldots\delta_{M}^{\bar{\jmath}_{\sigma''_{r,n}}},\\
\delta_{i_{\sigma''_{r,{u+1}}}}^{\bar{J}}\ldots\delta^{\bar{\jmath}_{\sigma''_{r,n}}}_{i_{\sigma_{l,1}}}\ldots\delta_{I}^{\bar{\jmath}_{\sigma_{l,u}}} & = & 
\delta_N^{\bar{J}}\delta_L^{\bar{M}}\delta_I^{\bar{K}}\quad \delta^{\bar{N}}_{i_{\sigma''_{r,{u+1}}}}\ldots\delta_{M}^{\bar{\jmath}_{\sigma''_{r,n}}}\quad \delta_{i_{\sigma_{l,1}}}^{\bar{L}}\ldots\delta^{\bar{\jmath}_{\sigma_{l,u}}}_{K}.
\end{eqnarray}
As $P_{\sigma_{\pi_l}}=P_{\pi_l}$ and $P_{\sigma''_{\pi_r}}=P_{\pi_r}$, we can identify two scalar subcurrents in Eq.~(\ref{eq:CDUeqV})
\begin{eqnarray}
\begin{cal}J\end{cal}_{h_1}^{K\bar{L}}(\pi_l) & = & \sum_{\sigma \in S_u}\delta_{i_{\sigma_{l,1}}}^{\bar{L}}\ldots\delta^{\bar{\jmath}_{\sigma_{l,u}}}_{K}\;J_{h_1}(\sigma_{\pi_l}),\\
\begin{cal}J\end{cal}_{h_2}^{M\bar{N}}(\pi_r) & = & \sum_{\sigma'' \in S_{n-u}}\delta^{\bar{N}}_{i_{\sigma''_{r,{u+1}}}}\ldots\delta_{M}^{\bar{\jmath}_{\sigma''_{r,n}}} \;J_{h_2}(\sigma''_{\pi_r}),
\end{eqnarray}
and so the $u=v$ term reads
\begin{equation}
\big(\delta_L^{\bar{J}}\delta_N^{\bar{K}}\delta_I^{\bar{M}}\;V_{AG}^{h_1,h_2,ab}(P_{\pi_l},P_{\pi_r})+\delta_N^{\bar{J}}\delta_L^{\bar{M}}\delta_I^{\bar{K}}\;V_{AG}^{h_2,h_1,ab}(P_{\pi_r},P_{\pi_l})\big)\begin{cal}J\end{cal}_{h_1}^{K\bar{L}}(\pi_l)\begin{cal}J\end{cal}_{h_2}^{M\bar{N}}(\pi_r).
\end{equation}
We define the color-dressed $AG$-vertex as\footnote{As for the Berends-Giele case, the color indices of the vertices are suppressed.}
\begin{equation}
\begin{cal}V\end{cal}_{AG}^{h_1,h_2,ab}(P_{\pi_l},P_{\pi_r})=\delta_L^{\bar{J}}\delta_N^{\bar{K}}\delta_I^{\bar{M}}\;V_{AG}^{h_1,h_2,ab}(P_{\pi_l},P_{\pi_r})+\delta_N^{\bar{J}}\delta_L^{\bar{M}}\delta_I^{\bar{K}}\;V_{AG}^{h_2,h_1,ab}(P_{\pi_r},P_{\pi_l}),
\end{equation}
and so finally the $u=v$ term reads
\begin{equation}
\begin{cal}V\end{cal}_{AG}^{h_1,h_2,ab}(P_{\pi_l},P_{\pi_r})\begin{cal}J\end{cal}_{h_1}^{K\bar{L}}(\pi_l)\begin{cal}J\end{cal}_{h_2}^{M\bar{N}}(\pi_r).
\end{equation}
We now turn to the $u\neq v$ term in Eq.~(\ref{eq:SCOdouble}). This term has four contributions, corresponding to the right-handed and left-handed decompositions for each of the two terms appearing in Eq.~(\ref{eq:CDdoubleline}). We will show the color dressing of the right-handed decomposition of the first term in Eq.~(\ref{eq:CDdoubleline}) explicitly. The three remaining contributions can be obtained in a similar way. The right-handed contribution to the first term in Eq.~(\ref{eq:CDdoubleline}) gives
\begin{eqnarray}
 & & \sum_{\sigma \in S_u}\sum_{\sigma' \in S_{v-u}}\sum_{\sigma'' \in S_{n-v}} \sum_{w=u}^{v-1} V_{AAR}^{a'b',h,ab}(P_{\sigma_{\pi_l}},P_{\sigma'_{m,w+1},\sigma'_{m,v}},P_{\sigma''_{\pi_r}})\\
 & & \qquad\qquad \delta_{i_{\sigma_{l,1}}}^{\bar{J}}\ldots\delta_{i_{\sigma'_{m,u+1}}}^{\bar{\jmath}_{\sigma_{l,u}}}\ldots 
\delta_{i_{\sigma''_{r,{v+1}}}}^{\bar{\jmath}_{\sigma'_{m,{v}}}}\ldots\delta_{I}^{\bar{\jmath}_{\sigma''_{r,n}}}\, J_{uw}^{a'b'}(\sigma_{\pi_l},\sigma'_{\pi_m})\,J_h(\sigma''_{\pi_r}).\nonumber
\end{eqnarray}
The color factor is decomposed in the usual way
\begin{equation}
\delta_{i_{\sigma_{l,1}}}^{\bar{J}}\ldots\delta_{i_{\sigma'_{m,u+1}}}^{\bar{\jmath}_{\sigma_{l,u}}}\ldots 
\delta_{i_{\sigma''_{r,{v+1}}}}^{\bar{\jmath}_{\sigma'_{m,{v}}}}\ldots\delta_{I}^{\bar{\jmath}_{\sigma''_{r,n}}} = \delta_L^{\bar{J}}\delta_N^{\bar{K}}\delta_I^{\bar{M}}\; 
\delta_{i_{\sigma_{l,1}}}^{\bar{L}}\ldots\delta_{i_{\sigma'_{m,u+1}}}^{\bar{\jmath}_{\sigma_{l,u}}}\ldots 
\delta_{K}^{\bar{\jmath}_{\sigma'_{m,{v}}}}\; \delta_{i_{\sigma''_{r,{v+1}}}}^{\bar{N}}\ldots\delta_{M}^{\bar{\jmath}_{\sigma''_{r,n}}}.
\end{equation}
As $P_{\sigma_{\pi_l}}=P_{\pi_l}$ and $P_{\sigma''_{\pi_r}}=P_{\pi_r}$, we can immediately identify the color-dressed scalar subcurrent
\begin{equation}
\begin{cal}J\end{cal}_{h}^{M\bar{N}}(\pi_r)  =  \sum_{\sigma'' \in S_{n-v}}\delta_{i_{\sigma''_{r,v+1}}}^{\bar{N}}\ldots\delta^{\bar{\jmath}_{\sigma''_{r,n}}}_{M}\;J_{h}(\sigma''_{\pi_r}),
\end{equation}
and we are left with
\begin{eqnarray}
 & & \sum_{\sigma \in S_u}\sum_{\sigma' \in S_{v-u}} \sum_{w=u}^{v-1} \delta_L^{\bar{J}}\delta_N^{\bar{K}}\delta_I^{\bar{M}} V_{AAR}^{a'b',h,ab}(P_{\pi_l},P_{\sigma'_{m,w+1},\sigma'_{m,v}},P_{\pi_r})\\
 & &  \qquad\qquad \delta_{i_{\sigma_{l,1}}}^{\bar{L}}\ldots\delta_{i_{\sigma'_{m,u+1}}}^{\bar{\jmath}_{\sigma_{l,u}}}\ldots 
\delta_{K}^{\bar{\jmath}_{\sigma'_{m,{v}}}}\, J_{uw}^{a'b'}(\sigma_{\pi_l},\sigma'_{\pi_m})\,\begin{cal}J\end{cal}_h^{M\bar{N}}(\pi_r).\nonumber
\end{eqnarray}
We will now consider the contributions coming from $w=u$ and $w>u$ separately. For $w=u$ we find, with $P_{\sigma'_{m,u+1},\sigma'_{m,v}}=P_{\sigma'_{\pi_m}}=P_{\pi_m}$,
\begin{equation}
\sum_{\sigma \in S_u}\sum_{\sigma' \in S_{v-u}} \delta_L^{\bar{J}}\delta_N^{\bar{K}}\delta_I^{\bar{M}} V_{AAR}^{a'b',h,ab}(P_{\pi_l},P_{\pi_m},P_{\pi_r}) \delta_{i_{\sigma_{l,1}}}^{\bar{L}}\ldots\delta_{i_{\sigma'_{m,u+1}}}^{\bar{\jmath}_{\sigma_{l,u}}}\ldots 
\delta_{K}^{\bar{\jmath}_{\sigma'_{m,{v}}}}\, J_{uu}^{a'b'}(\sigma_{\pi_l},\sigma'_{\pi_m})\,\begin{cal}J\end{cal}_h^{M\bar{N}}(\pi_r).
\label{eq:CDUneqV1}
\end{equation}
Due to the Kronecker-deltas appearing in the definition of $V_{AAR}$ (See Table~\ref{tab:csw_buiding_blocks}), one has
\begin{equation} 
V_{AAR}^{a'b',h,ab}(P_{\pi_l},P_{\pi_m},P_{\pi_r})\,J_{(v-u)(v-u)}^{a'b'}(\sigma'_{\pi_m},\sigma_{\pi_l})=0,
\label{eq:AARvanish1}
\end{equation}
and so Eq.~(\ref{eq:CDUneqV1}) can be written
\begin{eqnarray}
 & & \sum_{\sigma \in S_u}\sum_{\sigma' \in S_{v-u}} \delta_L^{\bar{J}}\delta_N^{\bar{K}}\delta_I^{\bar{M}} V_{AAR}^{a'b',h,ab}(P_{\pi_l},P_{\pi_m},P_{\pi_r})\begin{cal}J\end{cal}_h^{M\bar{N}}(\pi_r)\\ 
  & & \Big(\delta_{i_{\sigma_{l,1}}}^{\bar{L}}\ldots\delta_{i_{\sigma'_{m,u+1}}}^{\bar{\jmath}_{\sigma_{l,u}}}\ldots 
\delta_{K}^{\bar{\jmath}_{\sigma'_{m,{v}}}}\, J_{uu}^{a'b'}(\sigma_{\pi_l},\sigma'_{\pi_m}) + \delta_{i_{\sigma'_{m,u+1}}}^{\bar{L}}\ldots\delta_{i_{\sigma_{l,1}}}^{\bar{\jmath}_{\sigma'_{m,v}}}\ldots 
\delta_{K}^{\bar{\jmath}_{\sigma_{l,{u}}}}\, J_{(v-u)(v-u)}^{a'b'}(\sigma'_{\pi_m},\sigma_{\pi_l})\Big),\nonumber
\end{eqnarray}
where the second term vanishes according to Eq.~(\ref{eq:AARvanish1}). This second term corresponds to the second term in Eq.~(\ref{eq:CDdoubleline})
\begin{eqnarray}
\begin{cal}J\end{cal}_{\pi_l\pi_m}^{ab}(\pi_l,\pi_m) & = & \sum_{\sigma \in S_u}\sum_{\sigma' \in S_{v-u}} 
\Big(\delta_{i_{\sigma_{l,1}}}^{\bar{L}}\ldots\delta_{i_{\sigma'_{m,u+1}}}^{\bar{\jmath}_{\sigma_{l,u}}}\ldots 
\delta_{K}^{\bar{\jmath}_{\sigma'_{m,{v}}}}\, J_{uu}^{a'b'}(\sigma_{\pi_l},\sigma'_{\pi_m})\\
 & & \qquad \qquad \qquad  + \delta_{i_{\sigma'_{m,u+1}}}^{\bar{L}}\ldots\delta_{i_{\sigma_{l,1}}}^{\bar{\jmath}_{\sigma'_{m,v}}}\ldots 
\delta_{K}^{\bar{\jmath}_{\sigma_{l,{u}}}}\, J_{(v-u)(v-u)}^{a'b'}(\sigma'_{\pi_m},\sigma_{\pi_l})\Big)\nonumber.
\end{eqnarray}
So for $w=u$, we get
\begin{equation}
\delta_L^{\bar{J}}\delta_N^{\bar{K}}\delta_I^{\bar{M}} V_{AAR}^{a'b',h,ab}(P_{\pi_l},P_{\pi_m},P_{\pi_r})\begin{cal}J\end{cal}_{\pi_l\pi_m}^{a'b',\,K\bar{L}}(\pi_l,\pi_m)\begin{cal}J\end{cal}_h^{M\bar{N}}(\pi_r).
\end{equation}
For $w>u$, we can rewrite the sums over $w$ and $\sigma'$ in a similar way as in the Berends-Giele case, Eq.~(\ref{eq:OP3V}),
\begin{equation}
\sum_{w=u+1}^{v-1}\ \sum_{\sigma' \in S_{v-u}}\rightarrow \sum_{\pi \in OP(\pi_m,2)}\ \sum_{\sigma' \in S_{w-u}} \ \sum_{\sigma '' \in S_{v-w}},
\end{equation}
where $OP(\pi_m,2)$ is the set of all ordered partitions of $\pi_m$ into 2 independent parts. This rearrangement gives, with $P_{\sigma''_{m,w+1},\sigma''_{m,v}}=P_{\sigma''_{\pi_2}}=P_{\pi_2}$,
\begin{eqnarray}
 & & \sum_{\pi \in OP(\pi_m,2)}\sum_{\sigma \in S_u}\sum_{\sigma' \in S_{w-u}}\sum_{\sigma '' \in S_{v-w}} \delta_L^{\bar{J}}\delta_N^{\bar{K}}\delta_I^{\bar{M}} V_{AAR}^{a'b',h,ab}(P_{\pi_l},P_{\pi_2},P_{\pi_r})\\
 & &  \qquad\qquad \delta_{i_{\sigma_{l,1}}}^{\bar{L}}\ldots\delta_{i_{\sigma'_{m,u+1}}}^{\bar{\jmath}_{\sigma_{l,u}}}\ldots 
\delta_{K}^{\bar{\jmath}_{\sigma'_{m,{v}}}}\, J_{uw}^{a'b'}(\sigma_{\pi_l},\sigma'_{\pi_1},\sigma''_{\pi_2})\,\begin{cal}J\end{cal}_h^{M\bar{N}}(\pi_r),\nonumber
\end{eqnarray}
where $\pi =(\pi_1,\pi_2)$ is an ordered partition of $\pi_m$ and $w=\#\pi_1$. Due to the Kronecker-deltas in the definition of $V_{AAR}$, we have in a similar manner as for Eq.~(\ref{eq:AARvanish1})
\begin{equation}
V_{AAR}^{a'b',h,ab}(P_{\pi_l},P_{\pi_1},P_{\pi_r})J_{(v-w)(v-u)}^{a'b'}(\sigma''_{\pi_2},\sigma'_{\pi_1},\sigma_{\pi_l})=0.
\end{equation}
Using this relation, we can identify a color-dressed double-line current in a similar way as for the $w=u$ contribution. The $w>u$ contribution reads
\begin{equation}
\sum_{\pi\in OP(\pi_m,2)}\delta_L^{\bar{J}}\delta_N^{\bar{K}}\delta_I^{\bar{M}} V_{AAR}^{a'b',h,ab}(P_{\pi_l},P_{\pi_2},P_{\pi_r})\;\begin{cal}J\end{cal}_{\pi_l\pi_2}^{a'b',K\bar{L}}(\pi_l,\pi_1,\pi_2)\begin{cal}J\end{cal}_h^{M\bar{N}}(\pi_r).
\end{equation}
Putting together the $w=u$ and $w>u$ terms, we find the contribution from the right-handed decomposition of the first factor in Eq.~(\ref{eq:CDdoubleline})
\begin{eqnarray}
 & & \delta_L^{\bar{J}}\delta_N^{\bar{K}}\delta_I^{\bar{M}} V_{AAR}^{a'b',h,ab}(P_{\pi_l},P_{\pi_m},P_{\pi_r})\;    \begin{cal}J\end{cal}_{\pi_l\pi_m}^{a'b',K\bar{L}}(\pi_l,\pi_m)\begin{cal}J\end{cal}_h^{M\bar{N}}(\pi_r) \\
 & & \qquad\qquad + \sum_{\pi\in OP(\pi_m,2)} \delta_L^{\bar{J}}\delta_N^{\bar{K}}\delta_I^{\bar{M}} V_{AAR}^{a'b',h,ab}(P_{\pi_l},P_{\pi_2},P_{\pi_r}) \begin{cal}J\end{cal}_{\pi_l\pi_2}^{a'b',K\bar{L}}(\pi_l,\pi_1,\pi_2)\begin{cal}J\end{cal}_h^{M\bar{N}}(\pi_r). \nonumber
\end{eqnarray}
Similar terms are obtained for the remaining three contributions. Adding up all the contributions, we find for the $u\neq v$ term
\begin{eqnarray}
 & & \begin{cal}V\end{cal}_{AA}^{a'b',h,ab}(P_{\pi_l},P_{\pi_m},P_{\pi_r})\;\begin{cal}J\end{cal}_{\pi_l\pi_m}^{a'b',K\bar{L}}(\pi_l,\pi_m)\;\begin{cal}J\end{cal}_h^{M\bar{N}}(\pi_r)\\
 & & \qquad
+\;\begin{cal}V\end{cal}_{AA}^{a'b',h,ab}(P_{\pi_r},P_{\pi_m},P_{\pi_l})
\;\begin{cal}J\end{cal}_{\pi_m\pi_r}^{a'b',K\bar{L}}(\pi_m,\pi_r)\;\begin{cal}J\end{cal}_h^{M\bar{N}}(\pi_l)\nonumber\\
 & & \qquad + \sum_{\pi\in OP(\pi_m,2)}\Bigg[
 \begin{cal}V\end{cal}_{AA}^{a'b',h,ab}(P_{\pi_l},P_{\pi_2},P_{\pi_r}) \;\begin{cal}J\end{cal}_{\pi_l\pi_2}^{a'b',K\bar{L}}(\pi_l,\pi_1,\pi_2)\;\begin{cal}J\end{cal}_h^{M\bar{N}}(\pi_r)\nonumber \\
 & & \qquad \qquad \qquad \qquad \qquad+ \begin{cal}V\end{cal}_{AA}^{a'b',h,ab}(P_{\pi_r},P_{\pi_2},P_{\pi_l}) \;\begin{cal}J\end{cal}_{\pi_r\pi_2}^{a'b',K\bar{L}}(\pi_r,\pi_1,\pi_2)\;\begin{cal}J\end{cal}_h^{M\bar{N}}(\pi_l)\Bigg], \nonumber
\end{eqnarray}
where the color-dressed $AA$-vertex is defined as
\begin{equation}
\begin{cal}V\end{cal}_{AA}^{a'b',h,ab}(P_{\pi_1},P_{\pi_2},P_{\pi_3})=\delta^{\bar{J}}_L\delta^{\bar{K}}_N\delta^{\bar{M}}_I\; V_{AAR}^{a'b',h,ab}(P_{\pi_1},P_{\pi_2},P_{\pi_3})+ \delta^{\bar{J}}_N\delta^{\bar{M}}_L\delta^{\bar{K}}_I\; V_{AAL}^{h,a'b',ab}(P_{\pi_3},P_{\pi_2},P_{\pi_1}).
\end{equation}
Finally the recursive relations for the double-line current read
\begin{eqnarray}
 & & 
\begin{cal}J\end{cal}_{\pi_l,\pi_r}^{ab,I\bar{J}}(\pi_l,\pi_m,\pi_r)=\delta_{uv}\;\begin{cal}V\end{cal}_{AG}^{h_1,h_2,ab}(P_{\pi_l},P_{\pi_r})\begin{cal}J\end{cal}_{h_1}^{K\bar{L}}(\pi_l)\;\begin{cal}J\end{cal}_{h_2}^{M\bar{N}}(\pi_r)\\
 & & \qquad + (1-\delta_{uv})\frac{1}{2}\Bigg[ \begin{cal}V\end{cal}_{AA}^{a'b',h,ab}(P_{\pi_l},P_{\pi_m},P_{\pi_r})\;\begin{cal}J\end{cal}_{\pi_l\pi_m}^{a'b',K\bar{L}}(\pi_l,\pi_m)\;\begin{cal}J\end{cal}_h^{M\bar{N}}(\pi_r)\nonumber \\
  & & \qquad \qquad\qquad\qquad +\sum_{\pi\in OP(\pi_m,2)}
 \begin{cal}V\end{cal}_{AA}^{a'b',h,ab}(P_{\pi_l},P_{\pi_2},P_{\pi_r}) \;\begin{cal}J\end{cal}_{\pi_l\pi_2}^{a'b',K\bar{L}}(\pi_l,\pi_1,\pi_2)\;\begin{cal}J\end{cal}_h^{M\bar{N}}(\pi_r) \nonumber \\
  & & \qquad\qquad\qquad\qquad+(\pi_l \leftrightarrow \pi_r) \Bigg], \nonumber
\end{eqnarray}
which is Eq.~(\ref{eq:CDCSW2}) stated in Section~\ref{sec:CDCSW}.

We now turn to the color dressing of the recursive relations for the scalar off-shell currents, and we use again the symmetric form of the recursive relations, Eq.~(\ref{eq:SCOgluon}). The color dressing of the pure gluon part is identical to the color dressing of the three-gluon vertex part for the Berends-Giele recursive relations, and it evaluates to
\begin{equation}
\sum_{\pi\in P(n,2)}\begin{cal}V\end{cal}_{GG}^{h_1,h_2,h}(P_{\pi_1},P_{\pi_2})\begin{cal}J\end{cal}_{h_1}^{K\bar{L}}(\pi_1)\begin{cal}J\end{cal}_{h_2}^{M\bar{N}}(\pi_2).
\end{equation}
The term in Eq.~(\ref{eq:SCOgluon}) involving a double-line current has contributions from both the right-handed and the left-handed decompositions. We will only show the color dressing for the right-handed contribution here. The left-handed contribution can be obtained in a similar way. The right-handed contribution reads\footnote{Recall that this term has no contribution from $k=1$.}
\begin{equation}
\sum_{\sigma\in S_n}\sum_{k=2}^{n-1}\sum_{u=1}^{k-1} \sum_{v=u}^{k-1} \delta_{i_{\sigma_1}}^{\bar{J}}\ldots\delta_{I}^{\bar{\jmath}_{\sigma_n}} V_{GAR}^{ab,h_{1},h}(P_{\sigma_1,\sigma_u},P_{\sigma_{v+1},\sigma_k},P_{\sigma_{k+1},\sigma_n})\,
J_{uv}^{ab}(\sigma_1,\ldots,\sigma_k)\,J_{h_1}(\sigma_{k+1},\ldots,\sigma_n).
\label{eq:CDgluon1}
\end{equation}
We consider the terms corresponding to $u=v$ and $u\neq v$ separately. The $u=v$ term in Eq.~(\ref{eq:CDgluon1}) reads
\begin{equation}
\sum_{\sigma\in S_n}\sum_{k=2}^{n-1}\sum_{u=1}^{k-1}  \delta_{i_{\sigma_1}}^{\bar{J}}\ldots\delta_{I}^{\bar{\jmath}_{\sigma_n}} V_{GAR}^{ab,h_{1},h}(P_{\sigma_1,\sigma_u},P_{\sigma_{u+1},\sigma_k},P_{\sigma_{k+1},\sigma_n})\,
J_{uu}^{ab}(\sigma_1,\ldots,\sigma_k)\,J_{h_1}(\sigma_{k+1},\ldots,\sigma_n).
\label{eq:CDgluon2}
\end{equation}
The color factor is decomposed in the usual way
\begin{equation}
\delta_{i_{\sigma_1}}^{\bar{J}}\ldots\delta_{I}^{\bar{\jmath}_{\sigma_n}} = \delta^{\bar{J}}_L\delta^{\bar{K}}_N\delta^{\bar{M}}_I\; 
\delta_{i_{\sigma_1}}^{\bar{L}}\ldots\delta_{K}^{\bar{\jmath}_{\sigma_k}}\;
\delta_{i_{\sigma_{k+1}}}^{\bar{N}}\ldots\delta_{M}^{\bar{\jmath}_{\sigma_n}}.
\end{equation}
The sums appearing in this expression can be rearranged as
\begin{equation}
\sum_{\sigma \in S_n}\sum_{k=2}^{n-1}\sum_{u=1}^{k-1} \rightarrow \sum_{\pi \in OP(n,3)} \sum_{\sigma \in S_{u}} \sum_{\sigma' \in S_{k-u}} \sum_{\sigma'' \in S_{n-k}},
\end{equation}
where on the right-hand side $u =\# \pi_1$ and $k=n-\#\pi_3$ and $OP(n,3)$ is the set of all ordered partitions of $\{1,2,\ldots,n\}$ into three independent parts. Since 
\begin{equation}
V_{GAR}^{ab,h_1,h}(P_{\pi_1},P_{\pi_2},P_{\pi_3})J_{(k-u)(k-u)}^{ab}(\pi_2,\pi_1) = 0,
\end{equation}
due to the Kronecker-deltas in the definition of $V_{GAR}$, the result of the color dressing of Eq.~(\ref{eq:CDgluon2}) is
\begin{equation}
\sum_{\pi\in OP(n,3)} \delta^{\bar{J}}_L\delta^{\bar{K}}_N\delta^{\bar{M}}_I V_{GAR}^{ab,h_1,h}(P_{\pi_1},P_{\pi_2},P_{\pi_3}) 
\begin{cal}J\end{cal}_{\pi_1\pi_2}^{ab,K\bar{L}}(\pi_1,\pi_2)\begin{cal}J\end{cal}_{h_1}^{M\bar{N}}(\pi_3).
\end{equation}
Adding the left-handed contribution, we find
\begin{equation}
\sum_{\pi\in OP(n,3)} \begin{cal}V\end{cal}_{GA}^{ab,h_1,h}(P_{\pi_1},P_{\pi_2},P_{\pi_3}) 
\begin{cal}J\end{cal}_{\pi_1\pi_2}^{ab,K\bar{L}}(\pi_1,\pi_2)\begin{cal}J\end{cal}_{h_1}^{M\bar{N}}(\pi_3),
\end{equation}
where the color-dressed $GA$-vertex is defined by
\begin{equation}
\begin{cal}V\end{cal}_{GA}^{ab,h_1,h}(P_{\pi_1},P_{\pi_2},P_{\pi_3}) = 
\delta^{\bar{J}}_L\delta^{\bar{K}}_N\delta^{\bar{M}}_I V_{GAR}^{ab,h_1,h}(P_{\pi_1},P_{\pi_2},P_{\pi_3}) + 
\delta^{\bar{J}}_N\delta^{\bar{M}}_L\delta^{\bar{K}}_I V_{GAL}^{h_1,ab,h}(P_{\pi_3},P_{\pi_1},P_{\pi_2}).
\end{equation}
The color dressing of the $u\neq v$ contribution in Eq.~(\ref{eq:CDgluon1}) is similar to the $u=v$ case, except that we have to introduce the set $OP(n,4)$ of all ordered partitions of $\{1,\ldots,n\}$ into four independent parts, and the rearrangement of the sums is now written
\begin{equation}
\sum_{\sigma\in S_n}\sum_{k=1}^{n-1}\sum_{u=1}^{k-1} \sum_{v=u+1}^{k-1} \rightarrow
\sum_{\pi \in OP(n,4)}\sum_{\sigma \in S_u}\sum_{\sigma' \in S_{v-u}}\sum_{\sigma'' \in S_{k-v}}\sum_{\sigma''' \in S_{n-k}},
\end{equation}
where on the right hand side $u=\#\pi_1$, $v= \#\pi_1 + \#\pi_2$ and $k=n-\#\pi_4$. The result of the color dressing of the $u\neq v$ contribution then reads
\begin{equation}
\sum_{\pi\in OP(n,4)} \begin{cal}V\end{cal}_{GA}^{ab,h_1,h}(P_{\pi_1},P_{\pi_3},P_{\pi_4}) 
\begin{cal}J\end{cal}_{\pi_1\pi_3}^{ab,K\bar{L}}(\pi_1,\pi_2,\pi_3)\begin{cal}J\end{cal}_{h_1}^{M\bar{N}}(\pi_4).
\end{equation}
Finally the color-dressed recursive relations for the scalar off-shell current read
\begin{eqnarray}
\begin{cal}J\end{cal}_h^{I\bar{J}}(1,\ldots,n)=\frac{1}{P_{1,n}^2} & \Bigg[ & \sum_{\pi\in P(n,2)}\begin{cal}V\end{cal}_{GG}^{h_1,h_2,h}(P_{\pi_1},P_{\pi_2})\begin{cal}J\end{cal}_{h_1}^{K\bar{L}}(\pi_1)\begin{cal}J\end{cal}_{h_2}^{M\bar{N}}(\pi_2)\\
 & + & \frac{1}{2}\sum_{\pi\in OP(n,3)} \begin{cal}V\end{cal}_{GA}^{ab,h_1,h}(P_{\pi_1},P_{\pi_2},P_{\pi_3}) 
\begin{cal}J\end{cal}_{\pi_1\pi_2}^{ab,K\bar{L}}(\pi_1,\pi_2)\begin{cal}J\end{cal}_{h_1}^{M\bar{N}}(\pi_3)\nonumber \\
 & + & \frac{1}{2}\sum_{\pi\in OP(n,4)} \begin{cal}V\end{cal}_{GA}^{ab,h_1,h}(P_{\pi_1},P_{\pi_3},P_{\pi_4}) 
\begin{cal}J\end{cal}_{\pi_1\pi_3}^{ab,K\bar{L}}(\pi_1,\pi_2,\pi_3)\begin{cal}J\end{cal}_{h_1}^{M\bar{N}}(\pi_4)\Bigg ], \nonumber
\end{eqnarray}
which is Eq.~(\ref{eq:CDCSW1}) stated in Section~\ref{sec:CDCSW}.

\bibliography{database}
\end{document}